\documentclass[preprint, authoryear,11pt,fleqn]{elsarticle}
 \pdfoutput=1

\usepackage{amssymb,amsfonts,amsmath}
\usepackage{graphicx}
\usepackage{a4wide}
\usepackage{pxfonts}
\usepackage{ulem} 
\usepackage{color}
\usepackage{multirow}

\newcommand{\beg}{\begin{eqnarray}}
\newcommand{\fin}{\end{eqnarray}}
\newcommand{\magn}{{\bf B}}

\newcommand{\velo}{{\bf u}}
\newcommand{\vels}{u_s}

\newdimen\boxfigwidth 

\def\bigbox{\begingroup
  \boxfigwidth=\hsize
  \advance\boxfigwidth by -2\fboxrule
  \advance\boxfigwidth by -2\fboxsep
  \setbox4=\vbox\bgroup\hsize\boxfigwidth
  \hrule height0pt width\boxfigwidth\smallskip%
  \linewidth=\boxfigwidth
}
\def\endbigbox{\smallskip\egroup\fbox{\box4}\endgroup}

\newcommand{\zv}{\boldsymbol{e}_z}

\renewcommand{\velo}{{\mathbf u}}
\newcommand{\velong}{{\mathbf u}^a}
\newcommand{\Ping}{\Pi^a}
\newcommand{\lehnert}{\mathrm{Le}}
\newcommand{\lundquist}{\mathrm{Lu}}
\newcommand{\elsasser}{\mathrm{\Lambda}}
\newcommand{\ekman}{\mathrm{E}}
\renewcommand{\magn}{{\mathbf B}}

\journal{Physics of the Earth and Planetary Interiors}

\begin{document}
\begin{frontmatter}

\title{Hydromagnetic quasi-geostrophic modes in rapidly rotating planetary cores}
\author[EC]{E. Canet}
\author[EC,CF]{C. C. Finlay}
\author[AF]{A. Fournier}
\address[EC]{Earth and Planetary Magnetism Group, Institut f\"ur Geophysik, ETH Z\"urich, Sonneggstrasse 5, 8092 Z\"urich, Switzerland}
\address[CF]{Division of Geomagnetism, DTU Space, National Space Institute, Techical University of Denmark, Electrovej, 2800 Kngs. Lyngby, Denmark}
\address[AF]{G\'eomagn\'etisme, Institut de Physique du Globe de Paris, Sorbonne
Paris Cit\'e, Universit\'e Paris Diderot, UMR 7154 CNRS, F-75005 Paris, France}
\begin{abstract}
 The core of a terrestrial-type planet consists of a spherical shell of rapidly rotating, electrically conducting, fluid.  
Such a body supports two distinct classes of quasi-geostrophic (QG) eigenmodes: fast, primarily hydrodynamic, inertial modes with period related to the rotation time scale and slow, primarily magnetic, magnetostrophic modes with much longer periods. Here, we investigate the properties of these hydromagnetic quasi-geostrophic modes as a function of non-dimensional parameters controlling the strength of the background magnetic field, the planetary rotation rate, and 
the amount of magnetic dissipation.   

We first present analytic solutions that illustrate the essential parameter dependences of the modes and provide a convenient benchmark for our numerical scheme. A comparison between known three-dimensional inertial modes in a sphere and our axially invariant QG modes shows encouraging agreement at low azimuthal wavenumbers, particularly for the slowest modes. The container geometry and background magnetic field structure are found to influence the radial structure of the modes, but not the  scaling of their frequency with the control parameters.   When the background magnetic field decreases toward the outer boundary in a spherical shell, QG modes tend to be compressed towards the outer boundary.  Including magnetic dissipation, we find a continuous transition from diffusionless slow magnetic modes into quasi-free decay magnetic modes.  During that transition (which is controlled by the magnitude of the Elsasser number), we find that slow magnetic modes weakly modified by diffusion exhibit a distinctive spiralling planform. When magnetic diffusion is significant (Elsasser number much smaller than unity), we find quasi-free decay slow magnetic modes whose decay time scale is comparable to, or shorter than, their oscillation time scale.  

Based on our analysis, we expect Mercury to be in a regime where the slow magnetic modes are of quasi-free decay type. Earth and possibly Ganymede, with their larger Elsasser numbers, may possess slow modes that are in the transition regime of weak diffusion, depending on the details of their poorly known internal magnetic fields.  Fast QG modes, that are almost unaffected by the background magnetic field, are expected in the cores of all three bodies.

\end{abstract}


\end{frontmatter}

\section{Introduction} 
Electrically conducting fluids permeated by magnetic fields support hydromagnetic waves \citep{a1942}.  These waves play an important dynamical role.  They are a means of transporting energy and momentum, and are the natural response of the system to forcing. The cores of terrestrial-type planets, consisting of approximately spherical shells of rapidly-rotating liquid metal, will support such waves provided they also possess an intrinsic magnetic field.  Hydromagnetic waves in planetary cores are strongly influenced by rapid rotation which leads to the existence of two classes of waves, one with a time scale comparable to that of the rotation and another class with time scale much longer than that of rotation \citep{l1954, b1964, h1966}.  The latter has been proposed as candidates for producing geomagnetic secular variation and may also contribute to the time variations of other planetary magnetic fields.  Since planetary cores are closed containers, it is more appropriate to discuss free oscillations (or eigenmodes) rather than  progressive waves in an unconfined system; we therefore use the terminology hydromagnetic modes hereinafter.

In this study we determine the hydromagnetic modes for a simple quasi-geostrophic (QG) model of planetary core dynamics, building on the pioneering studies of \cite{h1966}, \cite{m1967} and \cite{b1976}.  Although hydromagnetic modes are often cited as an important ingredient of planetary core dynamics, there have been surprisingly few studies of them following these early important investigations.  The QG model \citep{b1970,p2008, c2009} assumes that disturbances are to leading order invariant parallel to the rotation axis.  Studies of three-dimensional instabilities and waves in rapidly-rotating hydromagnetic systems have demonstrated the QG model is a useful approximation provided the period of the disturbance considered is slow compared to the rotation period \citep{z1994,g2011} but faster than the magnetic diffusion time scale \citep{s2012}. 

Our model includes the spherical shell geometry essential for studies of planetary cores, rather than relying on a local $\beta$-plane \citep{h1966} or annulus \citep{b1976} approximation or being restricted to a full sphere \citep{m1967}. Besides assuming it to be uniform \citep{h1966} or linearly increasing with radius \citep{m1967}, we also consider the possibility of an underlying azimuthal magnetic field that reaches a maximum value within the shell, in line with the profile found by Gillet et al. (2010) for the cylindrical radial magnetic field. 
Furthermore, we include realistic levels of magnetic dissipation in the plane perpendicular to the rotation axis where the smallest length scales of the field perturbation are likely to be located.  Global mode solutions rather than local solutions are obtained numerically.  The investigations presented here are complementary to recent three-dimensional investigations of hydromagnetic modes by \cite{s2010, s2012} that focused on so-called quasi-free decay modes, which are closely related to the magnetic decay modes of a stationary conductor.  In order to concentrate on the intrinsic mode properties, we do not explicitly consider the forcing of the hydromagnetic modes; they may be forced, for example, by convection \cite[e.g.][]{f1983,z1995}, shear, or magnetic field instability \cite[e.g.][]{a1972, z1994}.  Our primary motivation is to better understand the linear dynamics possible within more complete (forced, nonlinear) QG models of rapid core dynamics, currently under development with the aim of modelling geomagnetic secular variation \cite[e.g.][]{c2009}.

Laboratory experiments involving rapidly rotating liquid metals permeated by strong magnetic fields also provide compelling evidence for the existence of hydromagnetic modes \citep{s2008,n2010}.  The detailed interpretation of these experiments is still under discussion, but it seems that the observed wave-like disturbances are influenced both by the magnetic field and by the rotation of the fluid.  There have also been reports that hydromagnetic waves may be responsible for some azimuthal motions of flux patches in numerical geodynamo models \cite[see e.g.][]{k2003}.  However, both in laboratory experiments and in numerical geodynamo models it has proven difficult to single out the particular wave mechanism at work.  Here, by studying a  simpler (linear, unforced) problem, and by making the additional assumption that the disturbances are axially invariant, we are able to fully characterize the nature of the oscillations.

Two fundamental non-dimensional parameters are found to control the properties of the hydromagnetic modes in our QG model.  The first is the ratio of the rotation time scale over the time scale for Alfv\'en waves to cross the system, known as the Lehnert number, 
\begin{eqnarray}
\lehnert=\frac{B^\star}{\Omega \sqrt{\rho \mu_0} r_o}.
\label{Lehnert}
\end{eqnarray}
The second is the ratio between the magnetic diffusion time scale and the time scale of Alfv\'en waves , known as the Lundquist number
\begin{eqnarray}
\lundquist=\frac{r_o B^\star}{ \eta \sqrt{\rho \mu_0}}. 
\label{Lundquist}
\end{eqnarray}
Here the outer core radius $r_o$ has been taken as the length scale and $B^\star$ is the typical magnetic field intensity in the planetary core interior. $\rho$ is the density of the fluid, $\Omega$ is the angular rotation rate, $\mu_0$ is the magnetic permeability of free space, and $\eta= 1 / \sigma \mu_0$ is the magnetic diffusivity ($\sigma$ is the electrical conductivity).  Table~\ref{tab:Lehnert} summarizes estimated ranges of the Lehnert and Lundquist numbers for Mercury, Earth and Ganymede.  For Earth's core an upper estimate of the field strength comes from the recent study by \cite{g2010} while for Mercury and Ganymede we use an upper estimate of ten times the observed field value downward continued to the core-mantle boundary (as suggested by the ratio between the mean field strength inside the shell and the surface poloidal field strength typically found in numerical geodynamo models).

\begin{table}{}
\scriptsize
\centering
\setlength{\tabcolsep}{0.5cm}
\begin{tabular}{|l|c|c|c|} \hline
& Mercury & Earth & Ganymede \\\hline & & &  \\
core radius ($\mathrm{m}$)       & $1.85\ 10^6$ $^{(c)}$ ; $2.02\ 10^6$ $^{(i)}$& $3.48\ 10^6$  & $4.8\ 10^5$ $^{(a)}$ \\
Field estimate ($\mathrm{T}$)      & $\left[1.4\ 10^{-6 \ (a)} - 1.4\ 10^{-5 \ (b)} \right]$ & $\left[7.6\ 10^{-4 \ (a)} - 3\ 10^{-3 \ (e)}\right]$ & $\left[2.5\ 10^{-4 \ (a)}- 2.5\ 10^{-3 \ (b)}\right]$  \\
Rotation rate$^{(c,a)}$ ($\mathrm{s}^{-1}$)& $1.24\ 10^{-6}$ & $7.27\ 10^{-5}$ & $1.02\ 10^{-5}$   \\
core density ($\mathrm{kg}/\mathrm{m}^3$) & $6980^{(i)}$ ; $8200^{(c)}$  & $11000^{(d)}$ & $6000^{(d)}$ \\
Electrical resistivity ($\mu \Omega \cdot \mathrm{m}$) & $0.36^{(g)} ; 0.44^{(g)} ; 0.9^{(f)} ; 1.67^{(c)}$ &$0.6^{(f)} ; 0.7^{(f)} ; 0.8^{(f)} ; 1.88^{(h)} $ & $0.8^{(f)} -5^{(d)}$ \\
Magnetic diffusivity ($\mathrm{m}^2/\mathrm{s}$) & $\left[0.29-1.3\right]$ & $\left[0.48 - 1.5\right]$ & $\left[0.64-4\right]$ \\ \hline & & &  \\
Ekman number $\ekman$ & $2\ 10^{-13}$ &  $10^{-15}$  & $4\ 10^{-13}$\\
Lehnert number $\lehnert$ & $\left[6\ 10^{-6} - 9\ 10^{-3}\right]$ & $ \left[2.5\ 10^{-5} - 10^{-4}\right]$ & $\left[5.8\ 10^{-4} - 5.8\ 10^{-3}\right]$ \\
Lundquist number $\lundquist$ & $\left[20-960\right]$ & $\left[14,750 - 186,000\right]$ & $\left[ 345-21,700 \right]$ \\
Elsasser number $\elsasser = \lehnert \ \lundquist$ & $\left[1.2\ 10^{-4} - 5.4 \ 10^{-2} \right]$  & $\left[0.4 - 18.6 \right]$ & $\left[0.2-125\right]$  \\
\hline
\end{tabular}
\caption{Estimates of core radius, magnetic field intensity, rotation rate, core density and magnetic diffusivity for the computation of Ekman, Lehnert, Lundquist and Elsasser numbers for Mercury, Earth, and Ganymede. The magnetic field lower estimates are taken from the maximum poloidal field downward continued from observations to the Core-Mantle boundary. The field upper estimates are estimates of the core interior field (from torsional waves studies for Earth, and using a factor 10 between the poloidal and toroidal field estimated from numerical dynamos for Mecury and Ganymede). These ranges of field values lead to the ranges for the Lehnert number.
For the computation of the magnetic diffusivities, we use a range of values of electrical resistivity including classical estimates from \cite{stevenson2003planetary,wicht2007,olson2007TOG} and recent estimates from first principles calculations of \cite{dekoker2012} (in agreement with \cite{pozzo2012}) or recent high pressure experiments of \cite{deng2013}. This leads to a range of possible Lundquist numbers.
\cite{deng2013} propose values at $7$ GPa and $1800-2200^\circ $C, for Mercury, we have taken De Koker et al.'s value at $7$ GPa and $2000^\circ $C, for Earth their values at $130$~GPa and $4000^\circ $C or $330$~ GPa and $5000^\circ $C for Fe, Fe-Si, Fe-O and for Ganymede $10$~GPa and $2000^\circ $C (pressure and temperature proposed by \cite{hauck2006} for a Fe-S composition).
For the computation of $\ekman=\nu / (\Omega r_0^2)$, following \cite{cebron2012elliptical}, we take a typical core kinematic viscosity $\nu = 10^{-6}$~m$^2$s$^{-1}$, consistent with a binary Fe/Fe-S composition.
Superscript $(a)$ for estimates from \cite{jones2003}, $(b)$ for ten times the observation based values, following numerical dynamo models. $(c)$ for \cite{wicht2007}, $(d)$ for \cite{stevenson2003planetary}, $(e)$ for \cite{g2010}, $(f)$ for \cite{dekoker2012}, $(g)$ for \cite{deng2013}, $(h)$ for \cite{olson2007TOG}, $(i)$ for \cite{hauck2013}. }
\label{tab:Lehnert}
\end{table}

We focus our discussion on the terrestrial-type planets since their interior most likely includes a spherical shell of liquid metal surrounded by a solid silicate layer, and because, as assumed in our model, incompressibility of the liquid layer is likely to be a reasonable approximation.  Since Venus and Mars do not presently possess an strong intrinisic field we discuss only Mercury, Earth and Jupiter's largest moon Ganymede, which also has a dynamo generated internal field.  Due to the lack of information concerning the aspect ratio of the inner core to outer core radii in Ganymede and Mercury, only the influence of changing $\lehnert$ and $\lundquist$ (both of which depend on $r_o$) is studied here, although the aspect ratio will certainly affect both the period and spatial structure of the eigenmodes.  Both the spatial and temporal spectrum of magnetic variations on Earth are now well known \citep{h2010}.  The nature of secular variation is currently unknown on Mercury \citep{a2008,a2012} and on Ganymede \citep{k2002}.  	Ambitious orbital missions carrying 
magnetometers are now planned to survey Mercury \cite[ESA's Bepi Colombo mission - see][]{y2004} and Ganymede \cite[ESA's Jupiter Icy Moon Explorer, JUICE, - see ][]{d2012} so it is of some interest to consider the role hydromagnetic QG modes could play in the magnetic secular variation of these bodies.  Crucially, Table~\ref{tab:Lehnert} shows that the Lehnert number is estimated to be very small for the cores of Mercury, Earth, and Ganymede, implying that rotation dominates magnetic forces for transient motions,  suggesting that a QG model is appropriate \citep{j2008} for studying them.  Estimates of the Lundquist number vary by several orders of magnitude, due to uncertainties both in the field strength and in the electrical resistivity of the planetary cores.  Nonetheless, it is certainly largest for the Earth and smallest for Mercury.  We will show that the Lunquist number plays a crucial role in determining the relevance of slow magnetic QG modes in planetary core dynamics.

This paper is divided into five sections.  Section \ref{sec:method} presents the mathematical formulation of our QG model.  Section \ref{sectAnalytic} presents an analytic solution describing many essential details of the modes including the splitting into fast and slow modes and the parameter dependence of the dispersion relations.  Detailed numerical results are presented in Section \ref{sec:results}.  These include the influence of spherical geometry, the influence of the background zonal field structure, and the effects of magnetic dissipation on the modes.  In Section  \ref{sec:discussion} the implications of our results for planetary core motions and variations in their magnetic fields are discussed.  An important comparison of modes periods in the absence of a magnetic field with the analytic full sphere inertial mode solutions is reported in \ref{app:zhang}.

\section{Model and methods}
\label{sec:method}

\subsection{Fluid flow}
\begin{figure}
\begin{minipage}{0.49\linewidth}
\centerline{\includegraphics[clip=true,width=\linewidth]{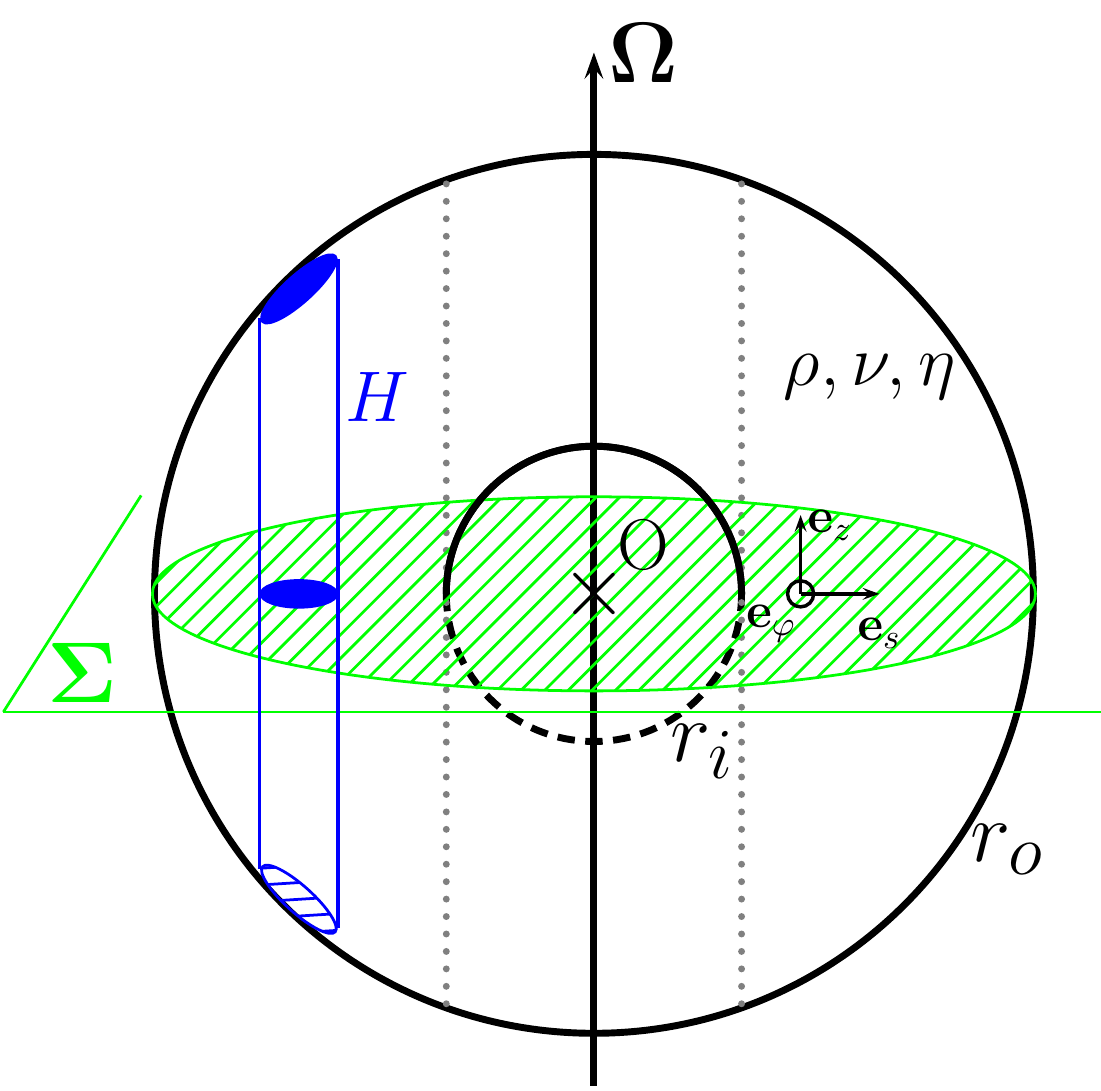}}
\caption{Side view of a planetary core. $\Sigma$ is the
equatorial plane and the core-mantle boundary is the outer 
 sphere, located at $r = r_o$. $r_i$ is the radius of the inner
core. 
$\rho, \nu, \eta$ define the fluid density, viscosity and magnetic diffusivity.
Dotted lines defines the cylinder tangent to the inner core. The column of height $2H$ illustrates an example columnar flow element (see text).}
\label{fig:geomsyst}
\end{minipage}
\end{figure}

As illustrated in Figure~\ref{fig:geomsyst}, we consider the dynamics of an
 incompressible fluid
contained in a spherical shell of inner and outer radii $r_i$ and $r_o$, 
respectively, assumed to have an Earth-like ratio $r_i/r_o=0.35$. The background planetary rotation 
 defines what we will refer to as the axial direction, 
 that is the $z$-axis in our working cylindrical 
 coordinate system $(s,\varphi,z)$. Note that we will restrict
 our attention to the dynamics occurring outside
 the tangent cylinder, where $r_i \leq s \leq r_o$, and that
 we will consider only non-zonal motions.  

 We are interested in the combined effect of 
 rotation and magnetic forces on the dynamical 
 behaviour of this system. 
 This behaviour is governed by the Navier--Stokes
 equation 
\begin{equation}
\rho \left( \frac{\mathrm{d}}{\mathrm{d}t} \velo 
          +2\boldsymbol{\Omega} 
          \times \velo \right)  
= 
- \boldsymbol{\nabla} \Pi + \rho \nu \nabla^2 \velo 
+ \frac{1}{\mu_0}{\mathbf B}\cdot{\boldsymbol \nabla}{\mathbf B}, 
\end{equation}
 in which  $\mathrm{d}/\mathrm{d}t$ designates the full material derivative, and $\velo$, $\Pi$ and $(1/\mu_0)
 {\mathbf B}\cdot{\boldsymbol \nabla}{\mathbf B}$ denote
 the fluid velocity, the modified pressure field and
 the magnetic tension, respectively (the magnetic pressure
 component of the Lorentz force has been absorbed into 
 the modified pressure field).
This equation has to be 
 supplemented by boundary conditions (which we will specify
 below), and is completed by the equation for mass conservation, assuming incompressibility,
\begin{equation}
\boldsymbol{\nabla} \cdot \velo = 0. 
\label{eq:massc}
\end{equation}
 The Lehnert number $\lehnert$ defined by (\ref{Lehnert}) above 
  quantifies the importance of rotation compared to magnetic effects
 in transient motions. 
 \cite{j2008} and \cite{g2011}
 showed in a series of numerical experiments that, if 
 $\lehnert \lesssim 10^{-2}$, transient motions possess 
 a substantial degree of axial invariance. Since the Lehnert
 number can be understood as the ratio of the time scale
 of inertial waves to the time scale of Alfv\'en waves, 
 small Lehnert number means that inertial waves react and communicate information from perturbations faster than Alfv\'en waves. This 
 enables a relaxation towards axial invariant geostrophy, despite the presence of magnetic forces. 
 The values of $\lehnert$ reported in Table~\ref{tab:Lehnert}
 for the Earth, Ganymede and Mercury are well
 below the $10^{-2}$ threshold; this motivates us to 
 introduce a quasi-geostrophic description of the fluid motions, by decomposing the flow $\velo$ and modified pressure $\Pi$ into a geostrophic
 component $(\velo^g,\Pi^g)$ and a non-geostrophic component $(\velong,\Ping)$, 
 according to 
 \begin{eqnarray}
 \velo &=& \velo^g + \velong,  \\ 
 \Pi &=& \Pi^g + \Ping. 
 \end{eqnarray}
 Our framework assumes that the main balance is
 geostrophic
 \begin{equation}
  2 \rho  \boldsymbol{\Omega} \times \velo^g = 
- \boldsymbol{\nabla} \Pi^g, 
 \label{eq:gb} 
 \end{equation}
 and that departures from this balance 
 are described by 
 \begin{equation}
\rho \left( \frac{\mathrm{d}}{\mathrm{d}t } \velo^g 
          +2\boldsymbol{\Omega}
                    \times \velong \right)  
= 
- \boldsymbol{\nabla} \Ping + \rho \nu \nabla^2 \velo^g 
 + \frac{1}{\mu_0}{\mathbf B}\cdot{\boldsymbol \nabla}{\mathbf B}.
 \label{eq:qgb}
 \end{equation}

 The quasi-geostrophic approximation therefore amounts to restricting
 the contribution of the non-geostrophic components of motion
 in this last equation to the Coriolis and pressure terms 
  \citep{g2011}. 
 The validity of this approximation 
  has been extensively discussed in the literature (consult
 for instance \cite{gillet2006qgmodel}, \cite{g2011}, 
 and references therein). 
 It suffices here to mention 
 that it stands on firm theoretical ground if the slope
 of the container, 
\begin{equation}
H'(s)\equiv{\mathrm d}H/{\mathrm d}s, \end{equation}
 remains small. 
  That condition does not hold in a spherical shell, for
 which 
\begin{equation}
H(s) = \sqrt{r_o^2 -s^2}. 
\end{equation}
 In practice however, the 
 quasi-geostrophic approximation can remain successful even
 if ${\mathrm d}H/{\mathrm d}s \sim 1$ \citep[see][]{williams2010}.

  It is now useful to consider the non-dimensional form taken by 
 (\ref{eq:qgb}), taking 
 $B^\star$ as the
 magnetic field scale,
the Alfv\'en wave speed
 \begin{equation}
 V_A \equiv \frac{B^\star}{\sqrt{\rho \mu_0}}
 \end{equation}
 as the velocity scale, the outer radius $r_o$ at the length scale, 
 the Alfv\'en time 
\begin{equation}
T_A \equiv \frac{r_o}{V_A} \label{eq:Ta}
\end{equation}
 as the time scale, and the kinetic pressure $\rho V_A^2$ as the pressure
 scale. 
This yields
\begin{equation}
\frac{\mathrm{d}}{\mathrm{d}t} \velo^g + \frac{2}{\lehnert} {\mathbf e}_z 
\times \velong = -\boldsymbol{\nabla}{\Ping} 
+ \frac{\ekman}{\lehnert} \nabla^2 \velo^g 
+ {\mathbf B}\cdot{\boldsymbol \nabla}{\mathbf B}
, 
\label{eq:nsnd}
\end{equation} 
where we understand that field variables are now non-dimensional. It
 appears 
 that viscous forces are proportional to the ratio
 of the Ekman number $\ekman$ to 
the Lehnert number $\lehnert$. The smallness of $\ekman$ 
 for the planetary cores of interest here is such that 
$\ekman/\lehnert \ll 1$ (see Table~\ref{tab:Lehnert}). We will therefore neglect viscous forces
 in the remainder of this study. 

 By virtue of the geostrophic balance written in (\ref{eq:gb}), the
 geostrophic flow does not vary along the axial direction. 
 We follow \cite{s2005} and \cite{c2009} and describe it
 as 
 \begin{equation}
 \velo^g (s,\varphi,t) = \frac{1}{H(s)} {\mathbf e}_z \times \left[
 \boldsymbol{\nabla}  \Psi (s,\varphi,t) 
\right],
 \label{eq:defpsi}
 \end{equation}  
 in which $\Psi$ is a time-dependent auxiliary streamfunction
 defined in the equatorial plane $\Sigma$ (recall Figure~\ref{fig:geomsyst}). 
 This definition, when inserted into \eqref{eq:massc}
 for three-dimensional
 mass conservation, implies a connection 
 between the $z$ component of $\velong$ 
 and the equatorial divergence of $\velo^g$, namely that
 \begin{equation}
 \boldsymbol{\nabla}_E \cdot \velo^g 
\equiv
 \frac{1}{s} 
 \left[ \partial_s \left( s u_s^g\right) + \partial_\varphi u_\varphi^g \right]= 
 -\beta u_s^g = -\partial_z u_z^a, 
 \label{eq:3dmass}
 \end{equation} 
 in which 
 \begin{equation}
 \beta(s) \equiv \frac{H'(s)}{H(s)}. 
 \label{eq:beta}
 \end{equation}
 The standard quasi-geostrophic approximation omits the $1/H(s)$ prefactor in (\ref{eq:defpsi}). 
 Consequently, it restricts the conservation of mass to the equatorial plane, and
 is only fully justified in the small slope situation discussed above. 
 In \ref{app:zhang}, we discuss further how faithfully this version of the quasi-geostrophic model can mimic fully three-dimensional (slow) inertial modes that have been analytically described by \cite{z2001}. 

  We now turn to the vorticity equation and consider the 
 $z$-component of the curl of the axial average of (\ref{eq:nsnd}) to obtain
 our working equation. It is expressed here directly in terms of $\Psi$, as
 \begin{equation}
\left[ 
 \frac{{\mathrm d}}{{\mathrm d}t} 
\left(
-\nabla^2_E + \beta \partial_s
\right)
- \frac{2}{\lehnert}  \frac{\beta}{s} \partial_\varphi \right] 
\Psi 
= H\, 
{\mathbf e}_z \cdot \boldsymbol{\nabla} \times 
\langle
\left(
{\mathbf B}\cdot{\boldsymbol \nabla}{\mathbf B}
\right)
\rangle,
 \label{eq:vort}
 \end{equation}
 where $\langle \cdot \rangle$ denotes axial averaging and 
 the equatorial Laplacian $\nabla^2_E $ is defined by 
\begin{equation}
\nabla^2_E \equiv \frac{1}{s} \partial_s s \partial_s + \frac{1}{s^2} \partial_\varphi^2.
\end{equation}
 Within the present inviscid framework, non-penetration at both boundaries 
 implies 
\begin{equation}
\Psi(s=r_i,\varphi,t) = \Psi(s=r_o,\varphi,t) =0 \ \ \forall (\varphi,t).  
\label{eq:bcf}
\end{equation}
 In addition, notice that our choice of $\Psi$ guarantees that
 $\velo \cdot {\mathbf e}_r = 0$ at $z \pm H(s), \ \forall (s,\varphi)$.\\ 

 As presented here, our description with the auxiliary function
 $\Psi$ does not account for zonal motions, which
 are beyond the scope of this study. Their consistent treatment would require
 the inclusion of the zonally averaged Navier--Stokes equation, 
 following for instance the analysis of \cite{plaut2002low}. 
 
\subsection{Magnetic induction}
 We have deferred until now the discussion of the evolution 
of the magnetic induction $\magn$, which is governed by
 the induction equation 
\begin{equation}
\partial_t \magn = \boldsymbol{\nabla} \times \left( \velo \times \magn \right)
+  \frac{1}{\mu_0 \sigma} \nabla^2 \magn, 
\end{equation}
under the constraint of conservation of magnetic flux
\begin{equation}
\boldsymbol{\nabla} \cdot \magn=0.
\end{equation}
The induction equation must be complemented by suitable boundary conditions (see below). 
 Its non-dimensional form is obtained using the scales defined above, which lead
 to 
\begin{equation}
\partial_t \magn = \boldsymbol{\nabla} \times \left( \velo \times \magn \right)
+ \frac{1}{\lundquist}\nabla^2 \magn, 
\end{equation}
where, again, we understand that
 field variables are non-dimensional. Magnetic
 diffusion scales in inverse proportion to the Lundquist number, previously
 defined in the introduction, and which expresses the ratio of the magnetic
 diffusion time scale to the Alfv\'en wave time scale $T_A$. This number
 is much larger than unity for the systems we are interested in, but the
 smallness of $1/\lundquist$ is by no means comparable to the smallness
 of the $\ekman/\lehnert$ ratio in the Navier--Stokes equation~(\ref{eq:nsnd}). 
 We therefore retain this term in our analysis; in fact, we 
 will show in the following that magnetic diffusion can 
 have dramatic consequences concerning the time-dependence and spatial structure of hydromagnetic QG modes. 

A consistent treatment of three-dimensional magnetic induction within the axially-averaged vorticity equation (\ref{eq:vort}) involves consideration of the evolution of three quadratic quantities that stem from the application of axial averaging to the magnetic tension  ${\mathbf B} \cdot \boldsymbol{\nabla}{\mathbf B}$, as described
 by \cite{c2009}. Here we instead make a major simplifying assumption, and consider
 only axially invariant magnetic fields, whose analysis can conveniently be restricted to the equatorial plane $\Sigma$.  The motivation behind this step is to obtain physical insight concerning the interaction of the magnetic field and rapidly rotating flow within a simple setting.  Furthermore when magnetic diffusion is a secondary effect, one expects perturbations of the magnetic field to follow those of the flow.  This assumption is not new; it was made by \cite{b1976} in his annulus model of the geodynamo,  and more recently by 
 \cite{tobias2007beta} in their $\beta$-plane
 analysis of MHD turbulence in the solar tachocline. 
 Furthermore, this assumption enables us to introduce consistently the effects of magnetic diffusion -- the quadratic formalism laid out by \cite{c2009} does not easily lend itself to including this important dissipative process.
 
 Our mathematical description of two-dimensional magnetic 
 induction $\magn$ is analogous to that of the flow, 
 and involves a magnetic potential $A$
 \begin{equation} 
 \magn(s,\varphi,t) = 
 \boldsymbol{\nabla} \times \left[
 A(s,\varphi,t) {\mathbf e}_z
 \right]. 
 \end{equation}
 This assumption allows us to rewrite the induction equation in the following scalar form
 \begin{equation}
 \partial_t A = -\velo^g \cdot \boldsymbol{\nabla} A 
              +  \frac{1}{\lundquist}\nabla^2_E
 A, 
 \label{eq:indA}
 \end{equation}
 in which  we notice that solely geostrophic flow 
 interacts with the magnetic potential. 
 In addition,  the magnetic tension term 
 ${\mathbf e}_z \cdot \boldsymbol{\nabla} \times 
(  {\mathbf B}\cdot{\boldsymbol \nabla}{\mathbf B})
$
 in (\ref{eq:vort}) now writes 
 $
 -\magn \cdot \boldsymbol{\nabla} (\nabla^2_E A)
 $. 
\subsection{Basic state and modal equations}
 The modal analysis of interest here requires that we linearize the
 problem around a basic state, by 
 decomposing 
\begin{eqnarray}
 \velo(\mathbf{r},t)&=&\mathbf{U}_0(\mathbf{r}) 
+ \delta\mathbf{\velo}(\mathbf{r},t),  \\
 \magn(\mathbf{r},t)&=&\magn_0(\mathbf{r}) 
+ \delta \magn(\mathbf{r},t). 
\end{eqnarray}
 We
 will assume throughout that the basic state is defined by a zero 
 velocity and a zonal magnetic field, whose strength 
 varies with cylindrical radius 
 \begin{eqnarray}
 \mathbf{U}_0 (\mathbf{r})&=& \mathbf{0}, \\
 \magn_0 (\mathbf{r})&=& B_0 (s) \mathbf{e}_\varphi.\
 \end{eqnarray}
 \begin{figure}
 \begin{minipage}{.49\linewidth}
\centerline{\includegraphics[clip=true,width=0.95\linewidth]{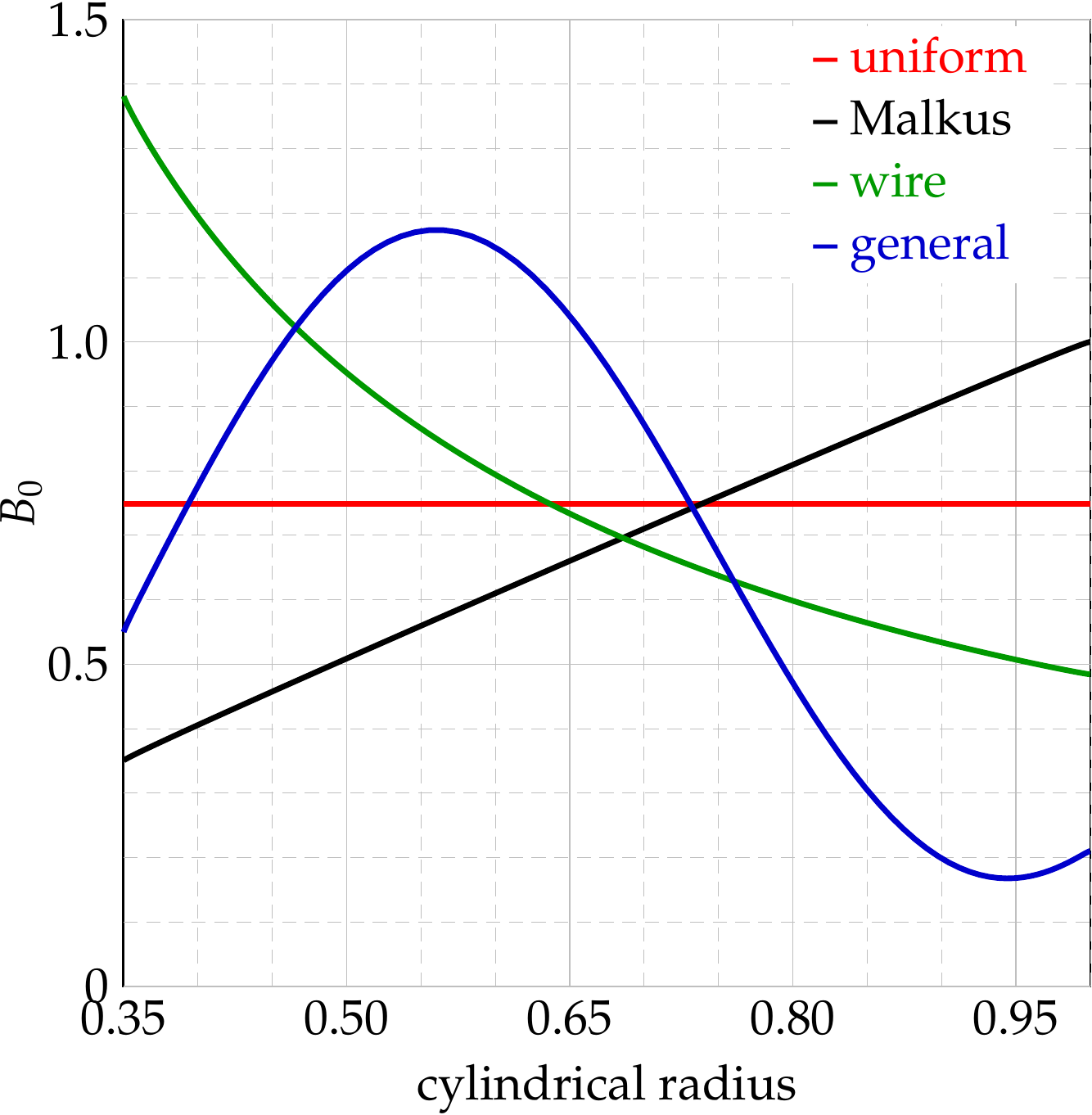}}
\caption{Radial profile of the zonal background fields $B_{0\varphi}$ studied. Red line: 
uniform toroidal field, black line: toroidal Malkus field (see text), green line: 
field corresponding to a wire, blue line: a more general zonal background field.}
\label{fig:magprof}
\end{minipage}
\end{figure}
The assumption of a purely zonal background magnetic field is a simplification; in reality the background field within planetary cores will also contain a (cylindrical) radial field which is neglected here to focus attention on the most essential physics. In order to study the impact of the background 
 magnetic field on the catalog of hydromagnetic modes, 
 we shall consider different background magnetic field profiles, as illustrated
 in Figure~\ref{fig:magprof}.
\renewcommand{\labelenumi}{(\roman{enumi})}
 \begin{enumerate}
 \item A uniform profile (red curve in Figure~\ref{fig:magprof}) 
       \begin{equation}
         B_0^u(s) = C_{un}, 
                  \label{eq:unif}
       \end{equation}
       with $C_{un}$ a constant. 
 \item The Malkus field, named after the
       seminal work of \citep{m1967} (black curve
       in Figure~\ref{fig:magprof}), which has
       \begin{equation}
         B_0^{\mathcal{M}}(s) = s. 
         \label{eq:malkusf}
       \end{equation}
 \item The field created by an electrical wire placed 
       along the $z$-axis (green curve in Figure~\ref{fig:magprof})
       \begin{equation}
         B_0^w(s) = C_w/s 
                  \label{eq:wiref}
       \end{equation}
       with $C_{w}$ a constant.
 \item A more general field, whose variations are non-monotonic
       (blue curve in Figure~\ref{fig:magprof}), and whose
       non-dimensional expression writes
      \begin{equation}
        B_0^g(s) = C_g f(s),    
                 \label{eq:gilletf}
      \end{equation}
      in which $C_g$ is a constant and 
     $f(s) = (3/2)\cos \left[ \pi(\alpha_1-\alpha_2s)\right]  +2$, with
      $\alpha_1=3/2$ and $\alpha_2=1/0.38$. 
 \end{enumerate}
 The constants $C_{un}, C_w$, and $C_g$  appearing in these expressions
 are chosen under the constraint of obtaining the same
 root-mean-square field strength in all cases, and
 therefore the same definition of the Lehnert number, independent
 of the background configuration. This leads to $C_{un}=0.75$, $C_w=0.48$ 
 and $C_g=0.34$.

 We proceed by utilizing the scalar functions $\Psi$ and $A$ introduced above in order 
 to describe the perturbations $\delta\mathbf{\velo}$
 and $\delta \magn$, respectively. In addition, let $A_0$ denote
 the magnetic potential which characterizes $\magn_0$. 
 Upon linearization, \eqref{eq:vort} 
 and \eqref{eq:indA} then become
 \begin{eqnarray}
 \left[ \partial_t \left( 
-\nabla^2_E + \beta \partial_s
\right)
- \frac{2}{\lehnert}  \frac{\beta}{s} \partial_\varphi 
 \right] \Psi &=&
 \frac{H}{s} \left( -{B_0} \partial_\varphi \nabla^2_E 
- \partial_s \nabla^2_E  A_0 \partial_\varphi
   \right)  A, \label{eq:pertf}\\
 \left( \partial_t  - \frac{1}{\lundquist}\nabla^2_E \right) A & =&  \frac{B_0}{sH} {\partial_\varphi} \Psi. \label{eq:perti}
 \end{eqnarray}
These are supplemented with non-penetration boundary 
 conditions for the flow (\ref{eq:bcf}), and vanishing magnetic perturbations at 
 both boundaries, following \cite{b1976}. This later condition writes
 \begin{equation}
A(s=r_i,\varphi,t) = A(s=r_o,\varphi,t) =0 \ \ \forall (\varphi,t).  
\label{eq:bci}
\end{equation}
We note in passing that setting $A=0$ at the boundaries is equivalent to assuming the boundaries are perfectly conducting. 

To carry out a global modal analysis, we consider trial 
 solutions of the form 
 \begin{equation}
 (\Psi,A) (s,\varphi,t) = \mathrm{Re}\left\{\left[\hat{\Psi}(s),\hat{A}(s)\right]
                                \exp \left[ -
                               \mathrm{i}
                                \left(
                               m \varphi + \omega t  
                                \right) 
                                \right] 
                                \right\},
 \end{equation}
where $m$ and $\omega$ are the azimuthal wavenumber and 
 angular frequency of the sought modes, respectively. 
Inserting this ansatz into
 \eqref{eq:pertf}-\eqref{eq:perti} leads to the following system
\begin{eqnarray}
 \left[ \omega \left( 
\frac{1}{s} \partial_s s \partial_s -\frac{m^2}{s^2} -\beta \partial_s 
\right) 
+ \frac{2}{\lehnert}\frac{m \beta}{s}
\right] \hat{\Psi} 
&=&
\frac{mH}{s} \left[
B_0 
\left(
\frac{1}{s}
 \partial_s s \partial_s -\frac{m^2}{s^2}
\right)
+\partial_s \nabla^2_E  A_0 
\right]
\hat{A}, \label{eq:psihat}\\ 
\left[\omega -\frac{\mathrm i}{\lundquist} \left(
\frac{1}{s}
 \partial_s s \partial_s -\frac{m^2}{s^2}
\right)
 \right] 
\hat{A} &=& \frac{m B_0}{sH} \hat{\Psi}. 
\label{eq:ahat}
\end{eqnarray}
We next approximate the cylindrical radial dependence of $\hat{\Psi}$
and $\hat{A}$ using Chebyshev polynomials \citep{trefethen2000spectral}. 
 If $\mathbf{x}$ denotes the state vector
 \begin{equation}
 \mathbf{x} \equiv \left[\hat{\Psi},\hat{A} \right]^T, 
 \end{equation} this collocation approach allows us to cast the modal problem
 at hand into of a generalized eigenvalue problem of the
 form   
 \begin{equation}
\mathbf{P}  \mathbf{x}= 
\omega \mathbf{Q}  \mathbf{x}
\label{eq:PQ}
 \end{equation}
in which $\mathbf{P}$ and $\mathbf{Q}$ are a complex-valued matrix
 and a real matrix, respectively. 
 We solve that problem using standard linear algebra software.

 To summarize,  let us stress that this problem is controlled by the background magnetic profile, the Lehnert number, and the Lundquist number. We will study in detail in section~\ref{sec:results} how each of these factors affects 
QG hydromagnetic modes in our spherical shell system.

\section{Analytic solution and benchmarking}
\label{sectAnalytic}

In order to gain more insight into the mode properties, 
we find it useful to first derive a QG analytic solution to equations~\eqref{eq:psihat}-\eqref{eq:ahat}. This derivation is akin to that 
proposed many years ago by \cite{m1967} in a full sphere, and it involves a background magnetic field increasing linearly with cylindrical radius as in (\ref{eq:malkusf}).
Furthermore, the solution requires that the small-slope approximation holds, namely that one can omit the $1/H(s)$ prefactor in~(\ref{eq:defpsi}), and correspondingly replace the operator
$\frac{1}{s} \partial_s s \partial_s -\frac{m^2}{s^2} -\beta \partial_s$ by 
$ \frac{1}{s} \partial_s s \partial_s -\frac{m^2}{s^2}$ 
in (\ref{eq:psihat}).\\ 

\subsection{Eigenfrequencies}

In the limit of small Lehnert number relevant to planetary cores,
 ($\lehnert \rightarrow 0$) and in the absence of dissipation ($\lundquist=+\infty$),
the Malkus field facilitates the derivation of analytic 
expressions for the frequency of the fast and slow modes, and allows 
their precise dependence on the Lehnert number $\lehnert$ to be determined. Under these conditions, we begin by combining \eqref{eq:psihat}-\eqref{eq:ahat} into 
a single equation for $\hat{\Psi}$
\begin{equation}
\left(\omega^2 - m^2 \right) 
\left( \frac{1}{s} \partial_s s \partial_s -\frac{m^2}{s^2} \right) \hat{\Psi}  + \omega   \frac{1}{\lehnert} \frac{2m\beta}{s} \hat{\Psi} = 0.
\label{eq:w1}
\end{equation}
Let us now introduce the frequency $\lambda$ as
\begin{equation}
\lambda\equiv\frac{\omega^2 - m^2}{\omega/\lehnert} \label{relationMR} 
\end{equation}
We can rearrange Equation~\eqref{eq:w1} into 
\begin{equation}
- \nabla^2_E \hat{\Psi}  = \frac{1}{\lambda} \frac{2m\beta}{s} \hat{\Psi}\label{eqmodes2D}. 
\end{equation}
Note that this last equation remains unchanged in the more general case of a magnetic perturbation
of the form $\delta \magn(\mathbf{r},t)= 
\boldsymbol{\nabla}\times \left[A(s,\varphi,t) \mathbf{e}_z\right] + b_z(s,\varphi,z,t) \mathbf{e}_z$, provided one defines $\lambda=(\omega^2 - m^2)/(\omega/\lehnert-m)$, a relation derived
 by \cite[][their Equation~27]{z2003} in their study of fully three-dimensional magneto-inertial waves. In our setting, the roots of (\ref{relationMR}) are 
 \begin{eqnarray}
 \omega_{\mathrm{f}} &=& \frac{\lambda}{2\lehnert} \left[1 + \sqrt{1+\frac{4 m^2 \lehnert^2 }{\lambda^2}}\right], \label{eq:wfe}\\
 \omega_{\mathrm{s}} &=& \frac{\lambda}{2\lehnert} \left[1 - \sqrt{1+\frac{4 m^2 \lehnert^2 }{\lambda^2}}\right]. \label{eq:wse} 
 \end{eqnarray}
 in which the subscripts $\mathrm{f}$ and $\mathrm{s}$ stand for fast and slow modes, respectively. 
In the limit $\lehnert\rightarrow 0$, these expressions can be further simplified
to first order as
\begin{eqnarray}
 \omega_{\mathrm{f}} &=& \frac{\lambda}{\lehnert}, \label{eq:wf} \\
 \omega_{\mathrm{s}} &=& -\frac{m^2\lehnert}{\lambda}. \label{eq:ws}
\end{eqnarray}
In terms of periods, this writes
\begin{eqnarray}
 T_{\mathrm{f}} &=& \frac{2 \pi \lehnert}{\lambda} , \label{eq:thTf}\\ 
 T_{\mathrm{s}} &=& -\frac{2 \pi \lambda}{m^2\lehnert}. \label{eq:thTs} 
\end{eqnarray}
The frequency of fast modes is essentially that of Rossby modes, and it is
of the order of the background angular velocity $\Omega$; the presence of the $1/\lehnert$ factor
 in \eqref{eq:wf} simply accounts 
for  the fact that the time scale we chose is the Alfv\'en wave time scale (\ref{eq:Ta}). Equation  (\ref{eqmodes2D}) is formally identical to the equation
obtained when seeking standard planetary Rossby modes, provided one sets $\lambda=\omega$. This implies in particular that the flow eigenfunctions are strictly the same for the hydrodynamic and hydromagnetic problems,  and in the latter case for both the fast and slow modes.

 The magnitude of the magnetic field eigenfunctions differs on the fast and slow branches. They are related through the diffusionless limit of \eqref{eq:ahat} -- in this section specified with the Malkus field and when the small-slope approximation holds --  by
 \begin{equation}
 \hat{A}_{\mathrm{f/s}} = \frac{m}{\omega_\mathrm{f/s}} \hat{\Psi}.  
 \end{equation}
 This expression allows us to compute the ratio of the kinetic energy $E_K$
 to the magnetic energy $E_M$ carried by a given mode. We find that
 \begin{equation}
 \left.\frac{E_M}{E_K}\right|_{\mathrm{f}/{\mathrm{s}}}=\frac{m^2}{\omega^2_{\mathrm{f}/\mathrm{s}}}
 \label{energy_ratio}
 \end{equation}
 showing that fast (resp. slow) modes essentially carry kinetic (resp. magnetic) energy, and
that in the planetary limit of low $\lehnert$ of interest in this study, equipartition between kinetic and magnetic energy characteristic of Alfv\'en waves is not relevant. 

\subsection{The constant $\beta$ case}
Equation \eqref{eqmodes2D} admits analytical solutions in the case of constant $\beta$. Since
$\beta=H'(s)/H(s)$, this corresponds to a container with 
an exponentially varying height $H(s)= e^{\beta s}$. We define the following change of variables 
\begin{eqnarray}
x&\equiv&\sqrt{s}, \\ 
\eta^2&\equiv&8m\beta/\lambda, 
\label{etalambda}
\end{eqnarray}
in which $\lambda$ is defined as in \eqref{relationMR}.  
Upon application of this change of variables, the wave equation \eqref{eqmodes2D} simplifies to
\begin{equation}
x\partial_x \left(x \partial_x \hat{\Psi} \right) + \left[\eta^2 x^2 - (2m)^2\right]
\hat{\Psi} = 0,
\end{equation}
in which we recognize a Bessel equation, whose solution writes  
\begin{equation}
\hat{\Psi}(s) = k_1 J_{2m} \left(\eta \sqrt{s}\right) 
+ k_2 Y_{2m} \left(\eta \sqrt{s}\right), 
\end{equation}
where 
$J_{2m}$ 
and 
$Y_{2m}$ 
are Bessel functions of order $2m$ of the first 
and second kind,  respectively. The constants $k_1$ and $k_2$
are determined by the boundary conditions \eqref{eq:bcf}, which
lead to
\begin{eqnarray}
k_1 J_{2m} (\eta \sqrt{r_i}) + k_2 Y_{2m} (\eta \sqrt{r_i})&=& 0, \label{det2}\\
k_1 J_{2m} (\eta \sqrt{r_o}) + k_2 Y_{2m} (\eta \sqrt{r_o})&=& 0. \label{determinant}
\end{eqnarray}
For this set to admit non-trivial solutions, its determinant $\mathcal{D}(\eta)$ must vanish. 
The zeroes of $\mathcal{D}$ define the collection of admissible $\eta$, and, by virtue of
 \eqref{etalambda} and \eqref{relationMR}, the corresponding eigenfrequencies. 
\begin{table}{}
\centering
\renewcommand{\arraystretch}{1.5} 
\setlength{\tabcolsep}{0.5cm}
\begin{tabular}{|l|c|c|c|c|} \hline
 & 1st zero & 2nd zero & 3rd zero & 4th zero\\\hline 
$\eta$  & $16.80$  &$21.88$ &$ 27.86$ & $34.50$ \\
$\lambda$  & $-0.17$ &$-0.10$ &$ -0.06$ & $-0.04$ \\
$\omega_\mathrm{f}$& $-1699.83$ &$-1002.71$ &$-618.49$ & $-380.95$\\
$\omega_\mathrm{s}$& $0.021 $&$ 0.036 $& $0.058$ & $0.095$ \\
$\left.E_M/E_K\right|_{\mathrm{f}}$ & $0.12\ 10^{-4} $ & $0.36\ 10^{-4}$ & $0.94\ 10^{-4}$ & $2.48\ 10^{-4}$ \\
$\left.E_M/E_K\right|_{\mathrm{s}}$&$8.03\ 10^{4}$   & $2.80\ 10^{4} $ & $1.06\ 10^{4}$ & $0.40\ 10^4$\\\hline 
\end{tabular}
\caption{
First four zeros of the determinant $\mathcal{D}$ of the system (\ref{det2}) and (\ref{determinant}), defined with a
background magnetic field increasingly linearly with radius and a constant radial rate
of change of the slope $\beta=-1$. Rows 2, 3 and 4 of this table
feature the corresponding 
auxiliary frequency $\lambda$ (defined by \eqref{etalambda}) and
the frequencies of the first four fast and slow modes, respectively, computed with $Le=10^{-4}$. 
Also shown for completeness are the ratios of magnetic to kinetic
energy carried by the modes.  All quantities are non-dimensional.}

\label{tablebetacst}
\end{table} 
As an illustration, Table \ref{tablebetacst} provides the first four 
zeros (in increasing order)
of $\mathcal{D}$ in the case $\beta=-1$ for the intermediate azimuthal wave number $m=6$. We see that frequency increases with mode
 number (higher radial modes have higher frequency), as in the canonical case of tension modes on a string attached at both ends. 
Note also that as in the string case, the $n$-th radial mode possesses $n+2$ nodes and  $n+1$ anti-nodes in range $[r_i,r_o]$.  
 Table~\ref{tablebetacst} also reveals than the non-dissipative slow modes of this setting oscillate on a time scale approximately one hundred times 
slower than the Alfv\'en wave time scale.

\begin{figure}
\begin{minipage}{0.49\linewidth}
\centerline{\includegraphics[clip=true,width=0.9\linewidth]{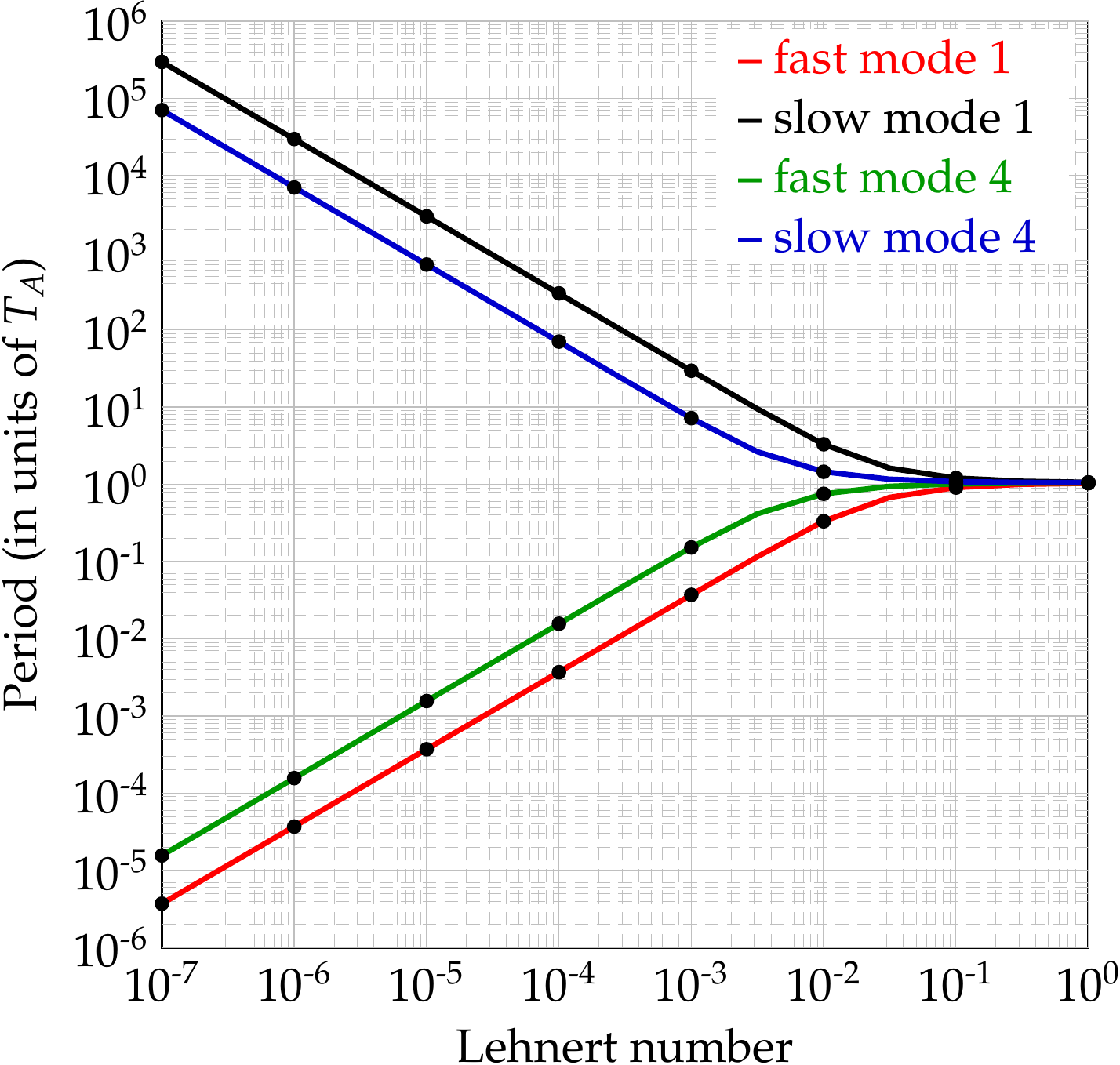}}
\end{minipage}
\begin{minipage}{0.49\linewidth}
\centerline{\includegraphics[clip=true,width=0.9\linewidth]{./figure3_leftpanel.pdf}}
\end{minipage}
\caption{Left: Period $T$ (expressed in Alfv\'en time) as a function of Lehnert number
$\lehnert$ for the first and fourth radial modes, for an azimuthal wavenumber $m=6$, in the 
same framework as reported in table~\ref{tablebetacst}. Right: Period $T$ as a function of azimuthal 
wavenumber $m$ for a fixed Lehnert number $\lehnert=10^{-4}$ for the first and fourth radial 
modes. Lines indicate theoretical periods, as taken from (44) and (45) in which 
$\lambda$ is also a function of $m$, black dots are calculated using our numerical code}
\label{analytic_results}
\end{figure}

Figure \ref{analytic_results} displays the periods of the modes in the  exponential container. The left panel shows the dependency of the period on the Lehnert number, for the first and fourth eigenmodes in radius, and for azimuthal wavenumber $m=6$. The results obtained in this global analytical case are in agreement with the asymptotic theory,
\eqref{eq:thTf}-\eqref{eq:thTs} above, as the period is proportional to $\lehnert$ on the fast branch and to $\lehnert^{-1}$ on the slow branch. The right panel represents the dependency of the period on $m$ at  $\lehnert=10^{-4}$, for the same eigenmodes. 
The periods seem to scale with $1/m$; this corresponds roughly to the zeros of the determinant $\mathcal{D}(\eta)$ since $m$ appears in the degree of the Bessel functions defining the determinant $\mathcal{D}$.  

The value of the above simple solution is that it provides exact values for $\lambda$ in~\eqref{relationMR}, which in turn allows the corresponding frequencies of the global modes to be exactly determined.  The corresponding modes furthermore possess a precise analytic form, depending in a relatively simple fashion on $\beta$.  These known eigenvalues and eigenfunctions provided us with a benchmark case that enabled us to validate the numerical method described in section \ref{sec:method} that was used in the following sections to compute catalogs of eigenmodes and eigenfrequencies for more general configurations of the boundary slope and the background zonal magnetic field.

\section{Numerical results}
\label{sec:results}
\subsection{Influence of container geometry: the hydrodynamic case}

We begin by numerically investigating the influence of the shape of the container on the spatial structure of the modes.  We compare the exponential container (also discussed 
in section \ref{sectAnalytic}, corresponding to a constant $\beta=-1$), a sloped cylinder (corresponding to a constant slope $H'(s)=-1$), and a spherical shell all of inner radius $r_i=0.35$ and outer radius $r_o=1$.  To simplify the comparison, we restrict the study to hydrodynamic Rossby modes defined by
\begin{equation}
 \left[ \omega \left( 
\frac{1}{s} \partial_s s \partial_s -\frac{m^2}{s^2} -\beta \partial_s 
\right) 
+ 2\frac{m \beta}{s}
\right] \hat{\Psi} 
=
0.
\label{eq:Rossby}
\end{equation}
Since the Alfv\'en time scale is meaningless in this non-magnetic context, time has here been scaled by the rotation time scale $\Omega^{-1}$ and velocities with $r_o \Omega$.  As in the hydromagnetic study, the information concerning the meridional shape of the axisymmetric container is contained in the $\beta$ parameter (first defined in Eq.~\eqref{eq:beta}). 
 
 \begin{figure}
 \begin{minipage}{.49\linewidth}
\centerline{\includegraphics[clip=true,width=0.95\linewidth]{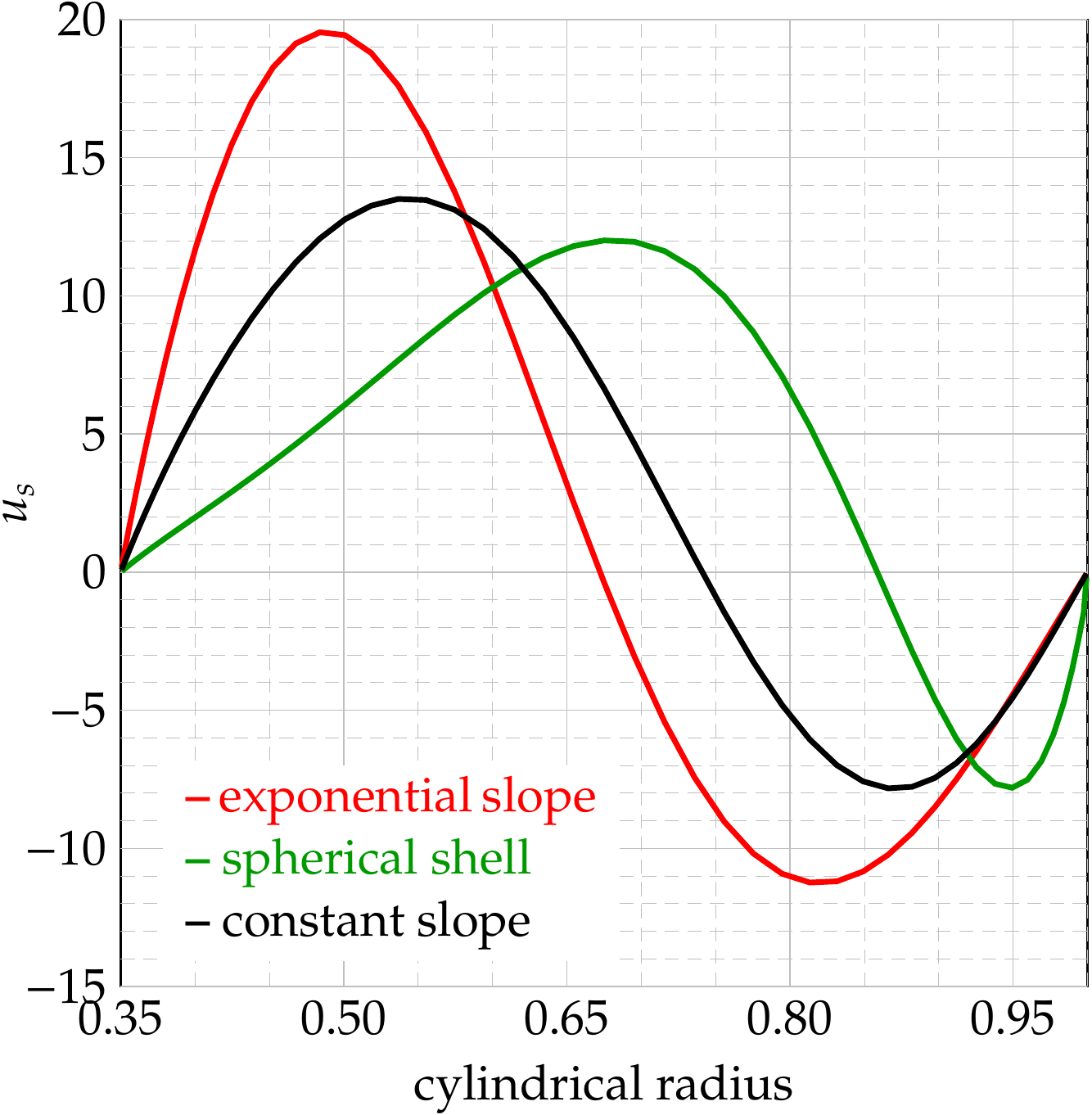}}
\caption{Influence of the shape of the container on the radial velocity ($\vels$) profile of the second eigenmode of azimuthal wavenumber $m=6$  as a function of the cylindrical radius $s$.
Black curve: sloped container (constant slope $H'(s)=-1$), red curve: exponential container (constant $\beta=-1$), green curve: spherical container}
\label{someQGmodes2}
\end{minipage}
\end{figure}

Figure \ref{someQGmodes2} displays radial velocity profiles corresponding to the second eigenmodes in radius for the azimuthal wavenumber $m=6$, as a function of the cylindrical radius for three different containers. 
For the spherical shell and the constant-slope annulus, $\beta$ and thus the restoring force depend on $s$; the restoring force is stronger further away
from the inner boundary -- since (i) the slope is larger (in the spherical case), or (ii) the column height is smaller (in the annulus) near the outer
boundary -- leading to a smaller node spacing in the vicinity of the outer boundary. For example, for the sixth eigenmode in a sphere, the non-dimensional distance between two nodes near the tangent cylinder is 0.2 whereas it drops to 0.03 near the outer boundary.  This results in the eigenfunction effectively being  concentrated towards
the outer boundary.

   The shape of the container therefore clearly does influence the form of global QG modes, and a simple constant $\beta$ theory cannot capture the full nature of modes in a spherical shell. 
 Another question of interest, in the spherical case, is the relevance of the QG approximation for accurately capturing global, fully three-dimensional inertial modes. We address this question in detail in \ref{app:zhang}. In a nutshell, we find a remarkable agreement (in terms of 
  the spectra predicted) when longest periods modes (with low $m$ and small radial wavenumber) are considered. 

\subsection{Influence of background zonal magnetic field}

\begin{figure}
Fundamental mode\\[2mm]
\begin{minipage}{0.49\linewidth}
$m=6$ \\[1mm]
\centerline{\includegraphics[clip=true,width=0.95\linewidth]{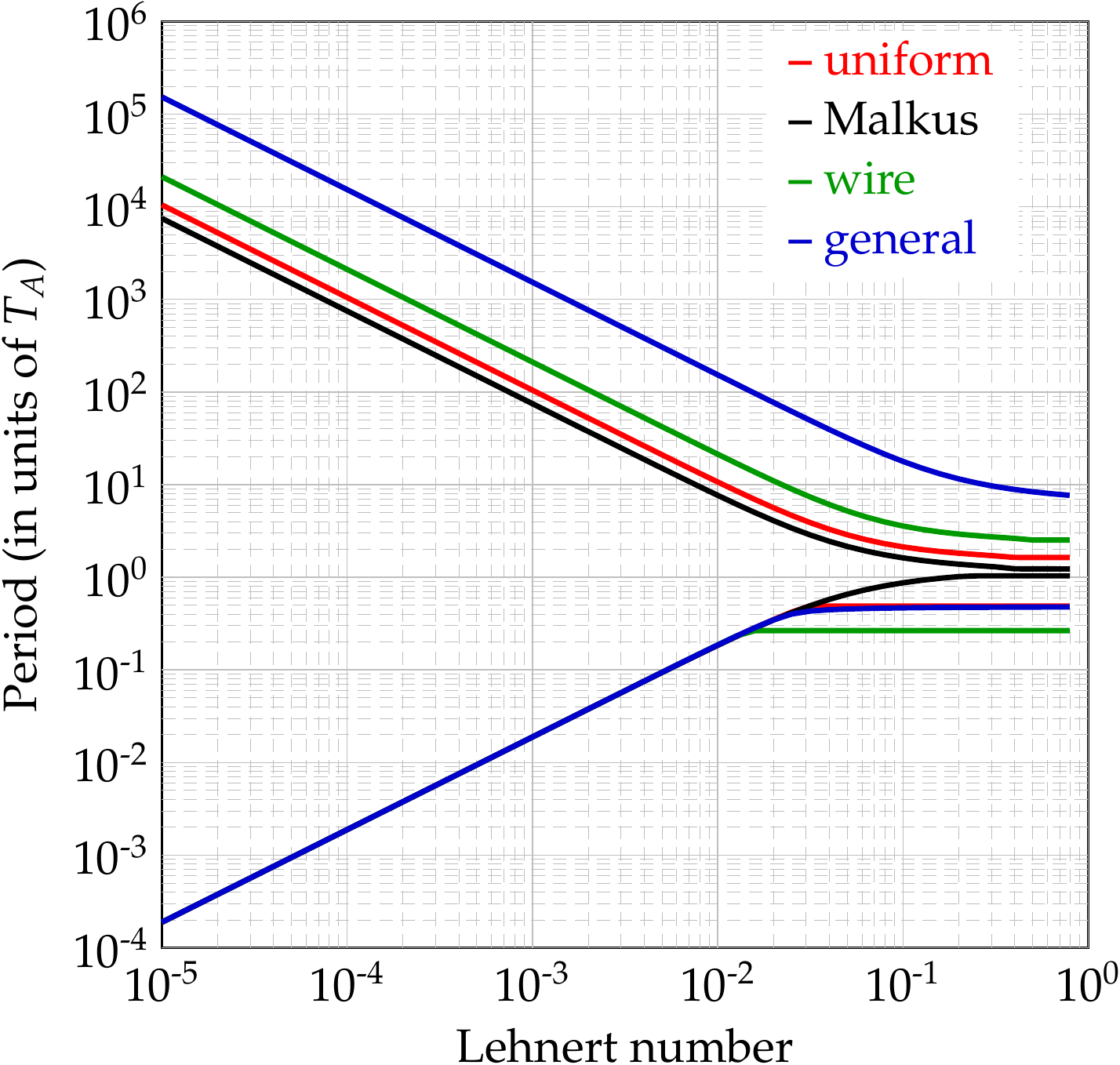}}
\end{minipage} 
\hfill
\begin{minipage}{0.49\linewidth}
$\lehnert=10^{-4}$ \\[1mm]
\centerline{\includegraphics[clip=true,width=0.95\linewidth]{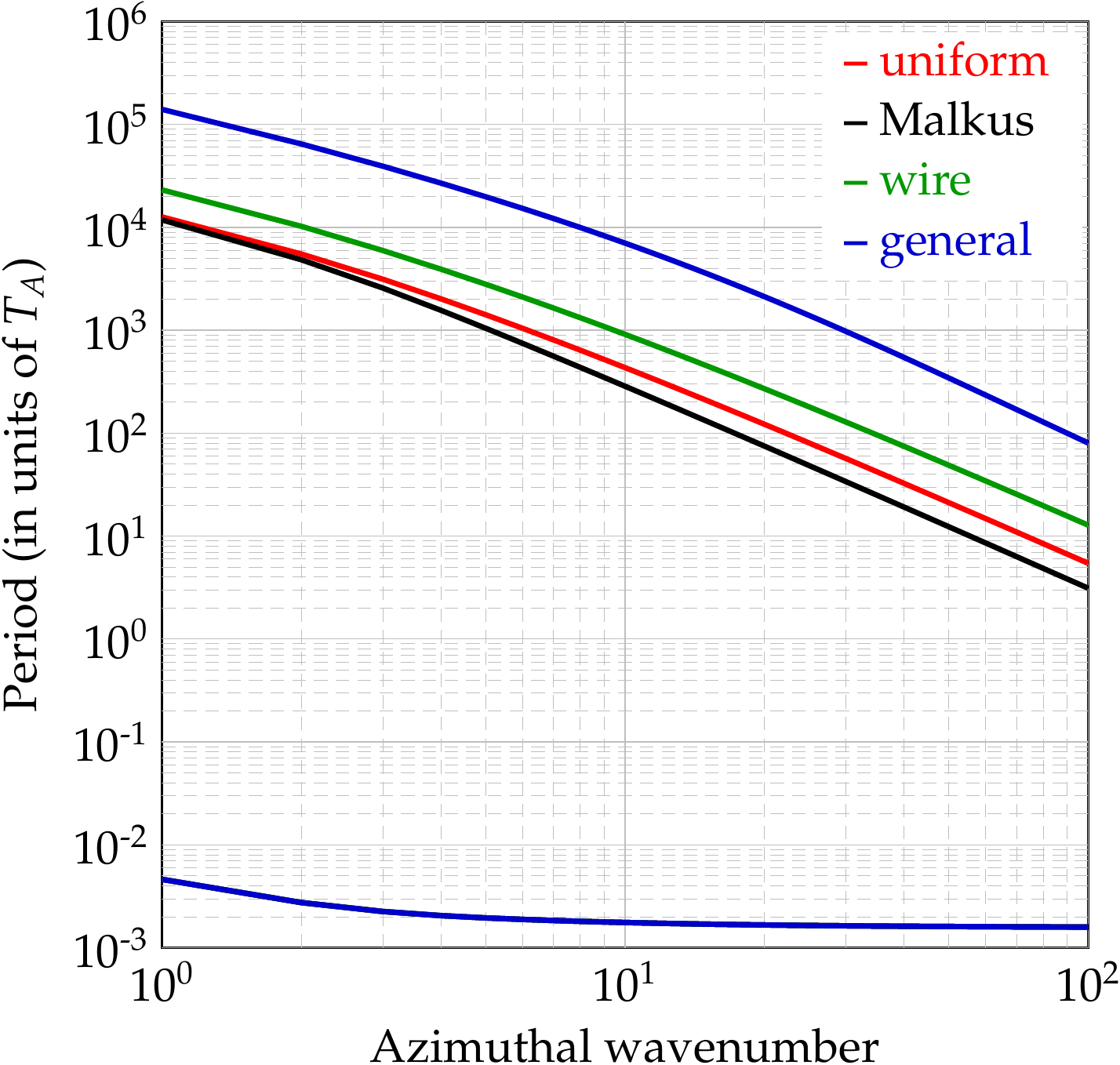}}
\end{minipage} 
\\[4mm]
Third overtone\\[2mm]
\begin{minipage}{0.49\linewidth}
$m=6$ \\[1mm]
\centerline{\includegraphics[clip=true,width=0.95\linewidth]{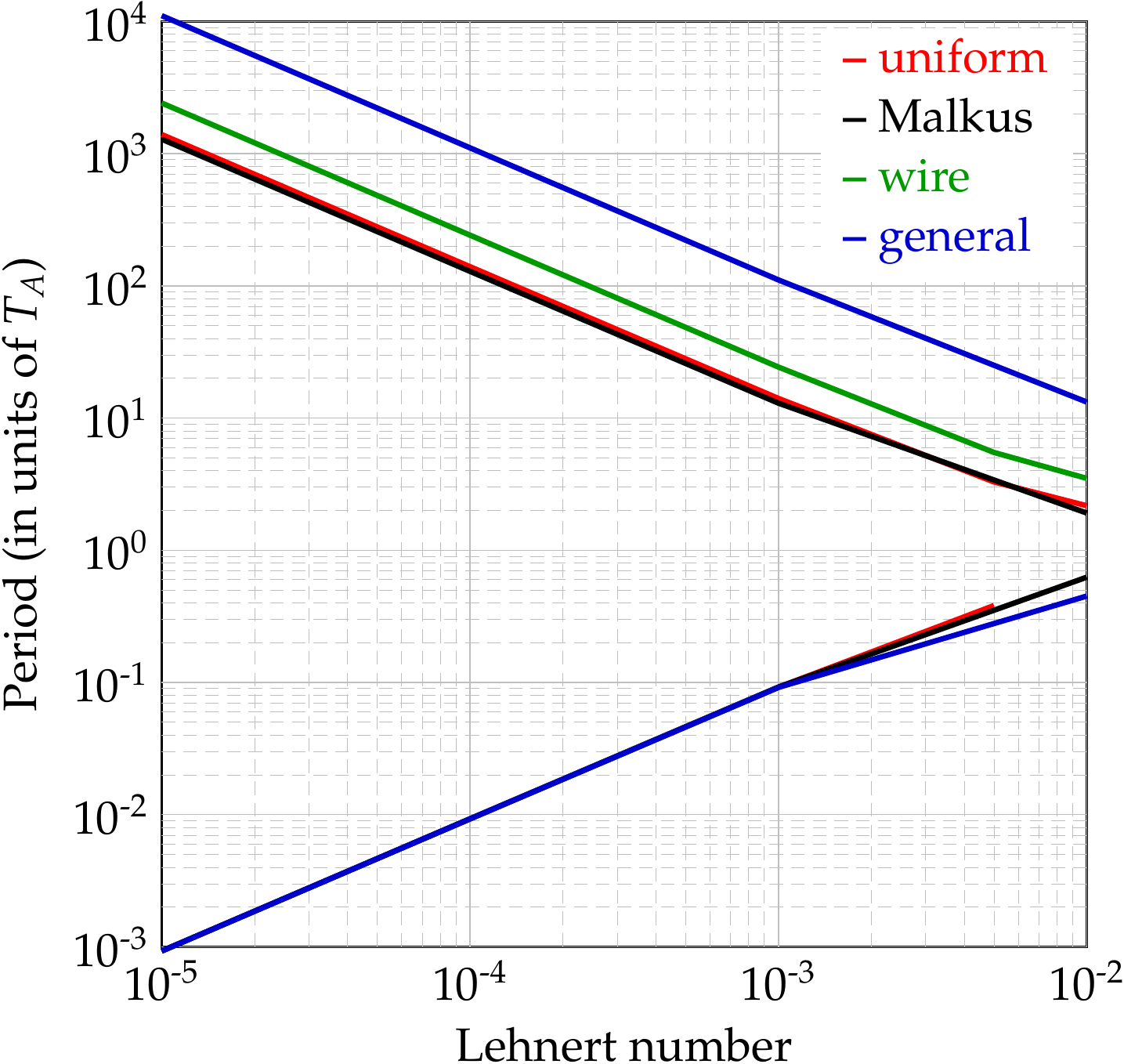}}
\end{minipage} 
\hfill
\begin{minipage}{0.49\linewidth}
$\lehnert=10^{-4}$ \\[1mm]
\centerline{\includegraphics[clip=true,width=0.95\linewidth]{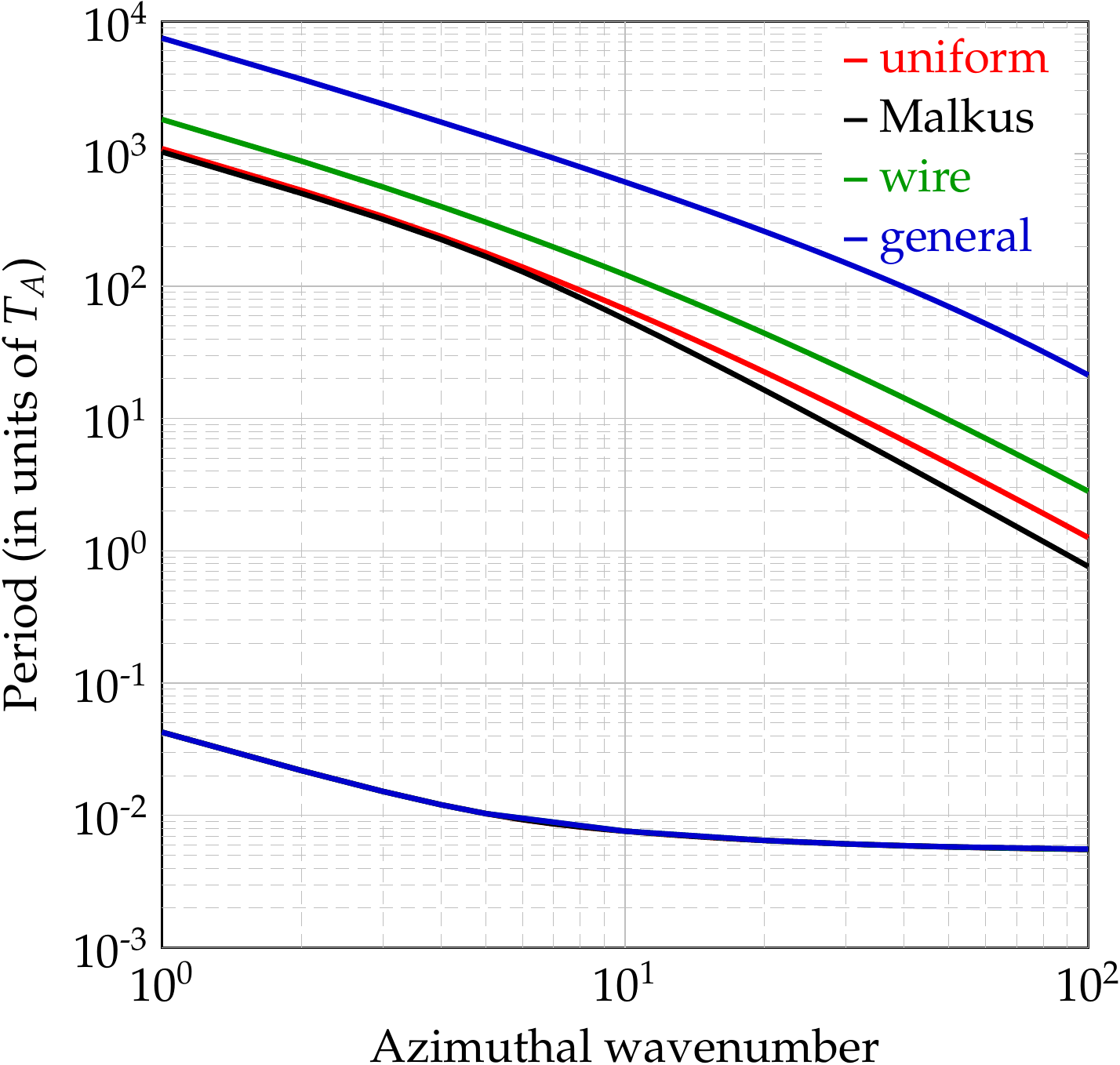}}
\end{minipage} 
\caption{Periods of the fast and slow QG modes for varying Lehnert number, azimuthal wavenumber, and
background magnetic field.  The top row corresponds to the fast and slow fundamental modes, and the bottom row to the fast and slow third overtones (fourth modes in the catalog). The periods shown in the left column are those of modes having azimuthal wavenumber $m=6$ (with varying Lehnert number and background field). Those shown in the right column have fixed Lehnert number ($\lehnert=10^{-4}$), and varying wavenumber
 and background field. The effect of four different background zonal magnetic fields is also presented: uniform field (red curves), Malkus field (black curves), field due to a wire at the axis of rotation (green curves) and a more general zonal field (blue curves).  Note that for the fast modes curves due to different fields overlap.}
\label{allfields1}
\end{figure}

These initial findings encourage us to return to the (more complete) hydromagnetic problem defined by Eqs.~\eqref{eq:psihat} and~\eqref{eq:ahat},  considering for now only the diffusionless scenario. We  numerically find solutions to the eigenvalue problem~\eqref{eq:PQ},  obtaining both flow and field eigenfunctions and their corresponding eigenfrequencies.  We explore the influence of different background zonal field profiles defined in \eqref{eq:unif} - \eqref{eq:gilletf} on the dispersion relation and spatial structure of the QG hydromagnetic modes in the spherical shell geometry most relevant for planetary cores.  Figure \ref{allfields1} displays the periods for the fast and slow modes as 
a function of $\lehnert$ (Left panels, with $m=6$) and a function of the azimuthal wavenumber $m$ (Right panels, with $\lehnert=10^{-4}$) for the four different background magnetic fields.  The upper and lower panels corresponds to the radial structure of the mode, namely the first (upper) and fourth (lower) modes in radius.

The background magnetic field structure does not influence the scaling of the mode period with respect to $\lehnert$ and $m$.  The scalings found here numerically for the slow waves are in agreement with the asymptotic theory presented earlier (see Eqs.~\eqref{eq:thTf} and ~\eqref{eq:thTs}).  The fast mode, essentially a Rossby mode, is not sensitive to the background magnetic field for $\lehnert<10^{-2}$. Its period is proportional to $\lehnert$ and depends only weakly on the azimuthal wavenumber.

\begin{figure}
\begin{minipage}{.49\linewidth}
\centerline{\includegraphics[clip=true,width=0.95\linewidth]{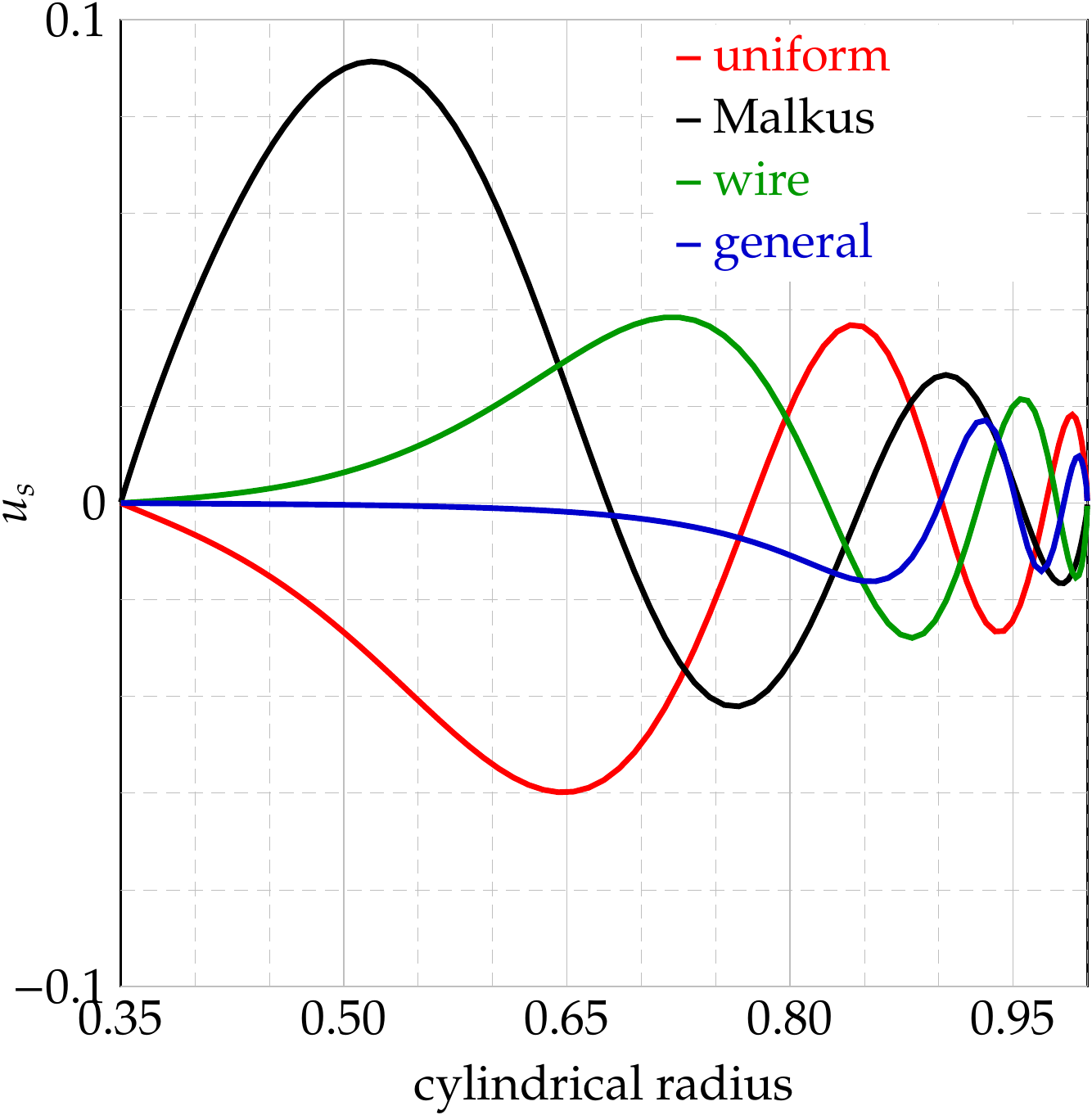}}
\caption{Profile of the radial velocity $\vels$ for the slow magnetic modes (4th modes (third overtones) in radius, $m=6$) for the four background magnetic fields described in the text: uniform field (red curve), Malkus field (black curve), field due to a wire current at the axis of rotation (green curve) and a more general zonal field (blue curve).}
\label{allfields2}
\end{minipage}
\end{figure}

In the limit of rapid rotation ($\lehnert<10^{-2}$), the period of the slow mode scales as the inverse of $\lehnert$ but with the modes due to different zonal background fields offset from one another.  The relation between the spatial structure of the background field and the exact period of the slow waves is not straigthforward.  For example, the non-monotonic zonal field (labelled general) is the most complicated in space, involving large spatial gradients  but the global analysis, perhaps surprisingly, shows that in this case, modes with longer periods are in fact produced, with the modes being concentrated near the outer boundary where the magnetic field is weakest (see Fig.~\ref{allfields2}).  The restoring Lorentz force most relevant to the eigenmodes arises due to the gradients generated by the time-varying, disturbance magnetic field. Because the general background field is weakest at large s, and lower than any other choice of background field considered, the eigenfunctions can condense into that region resulting in a longer period for the modes.

\subsection{Influence of Ohmic dissipation} 

As described above, the slow modes are essentially magnetic, while the fast modes are essentially hydrodynamic and only weakly feel the magnetic field.  When magnetic dissipation is included, local theory for plane hydromagnetic waves \citep[e.g.][section~2.5]{rudiger2005} suggests that
the fast wave, which is essentially of inertial type, is not affected by magnetic dissipation, 
whereas the slow one is essentially forced to decay.  How the structure of the mode is affected by magnetic diffusion is not addressed by the local theory.  Here, we solve problem~\eqref{eq:PQ} in spherical shell geometry, in the presence of the general zonal background magnetic field, and including magnetic dissipation whose level is controlled by the Lundquist number as defined in \eqref{Lundquist}.

\newcommand{\mylw}{.9\linewidth}
\begin{figure}
\begin{center}
\begin{minipage}{.9\linewidth}
\begin{minipage}{0.49\linewidth}
\centerline{$\Psi$, $\lundquist=+\infty$, slow mode}
\centerline{\includegraphics[clip=true,width=\mylw]{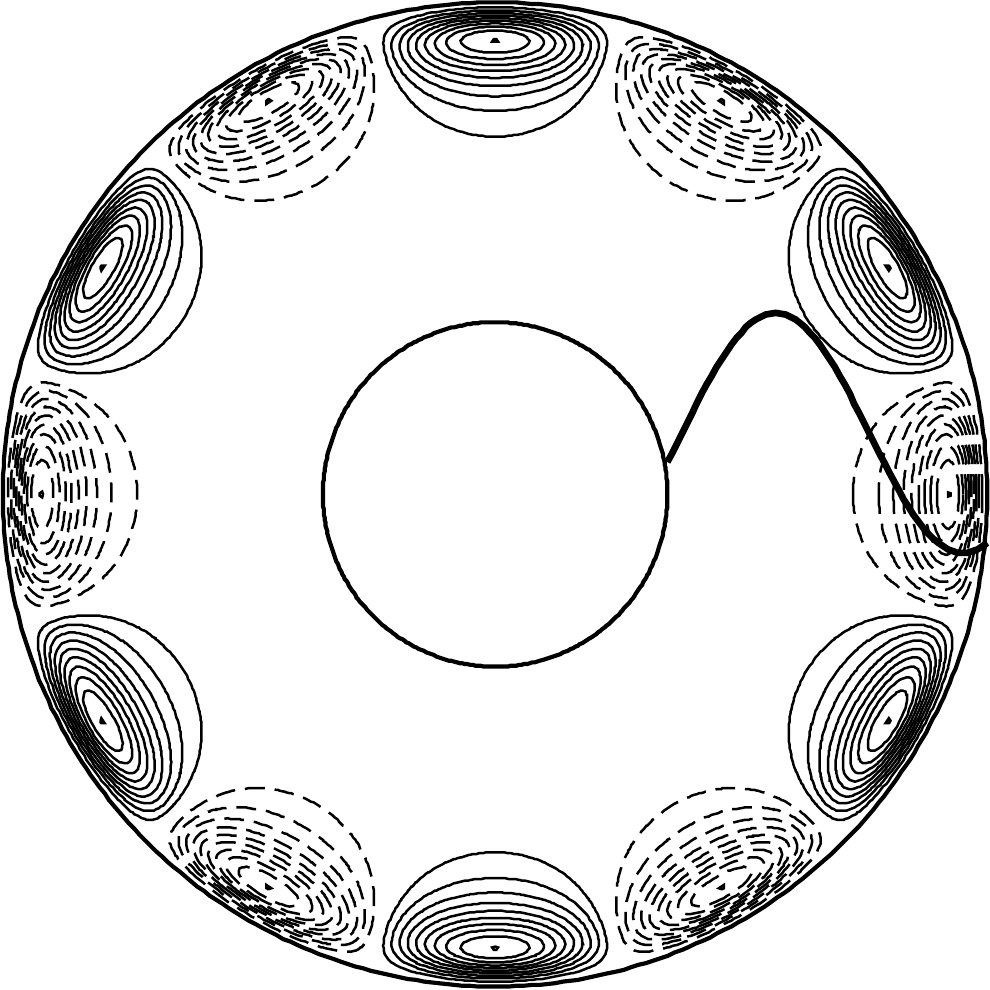}}
\end{minipage}
\hfill 
\begin{minipage}{0.49\linewidth}
\centerline{$\Psi$, $\lundquist=+\infty$, fast mode}
\centerline{\includegraphics[clip=true,width=\mylw]{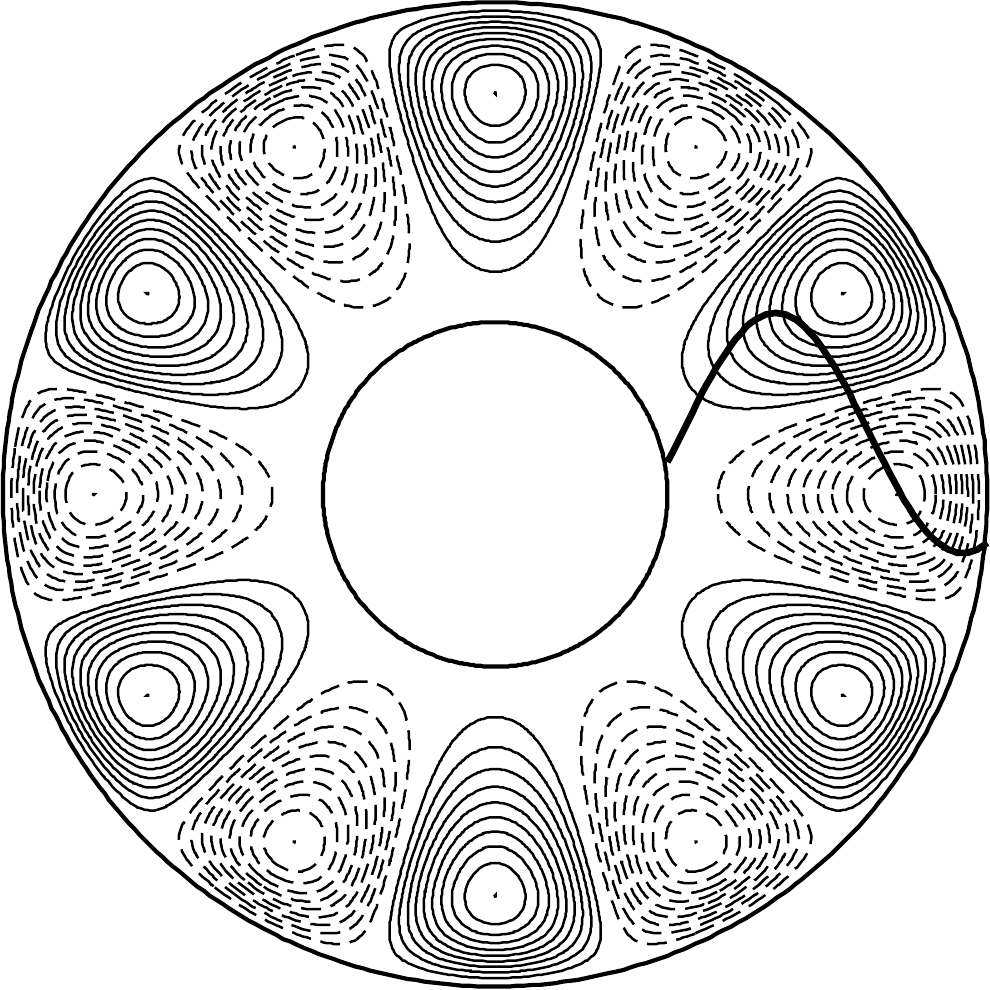}}
\end{minipage}
\\[2mm]
\begin{minipage}{0.49\linewidth}
\centerline{$\Psi$, $\lundquist=10^4$, slow mode}
\centerline{\includegraphics[clip=true,width=\mylw]{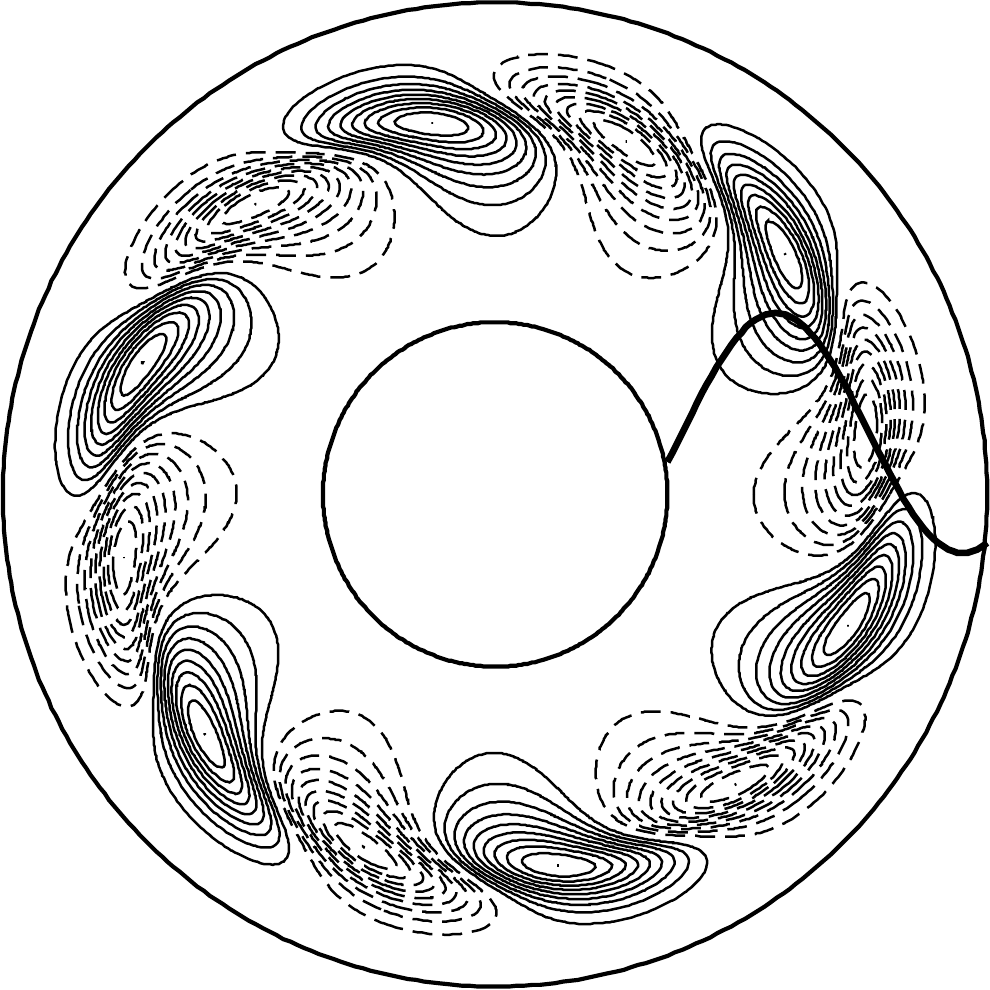}}
\end{minipage}
\hfill
\begin{minipage}{0.49\linewidth}
\centerline{$\Psi$, $\lundquist=10^4$, fast mode}
\centerline{\includegraphics[clip=true,width=\mylw]{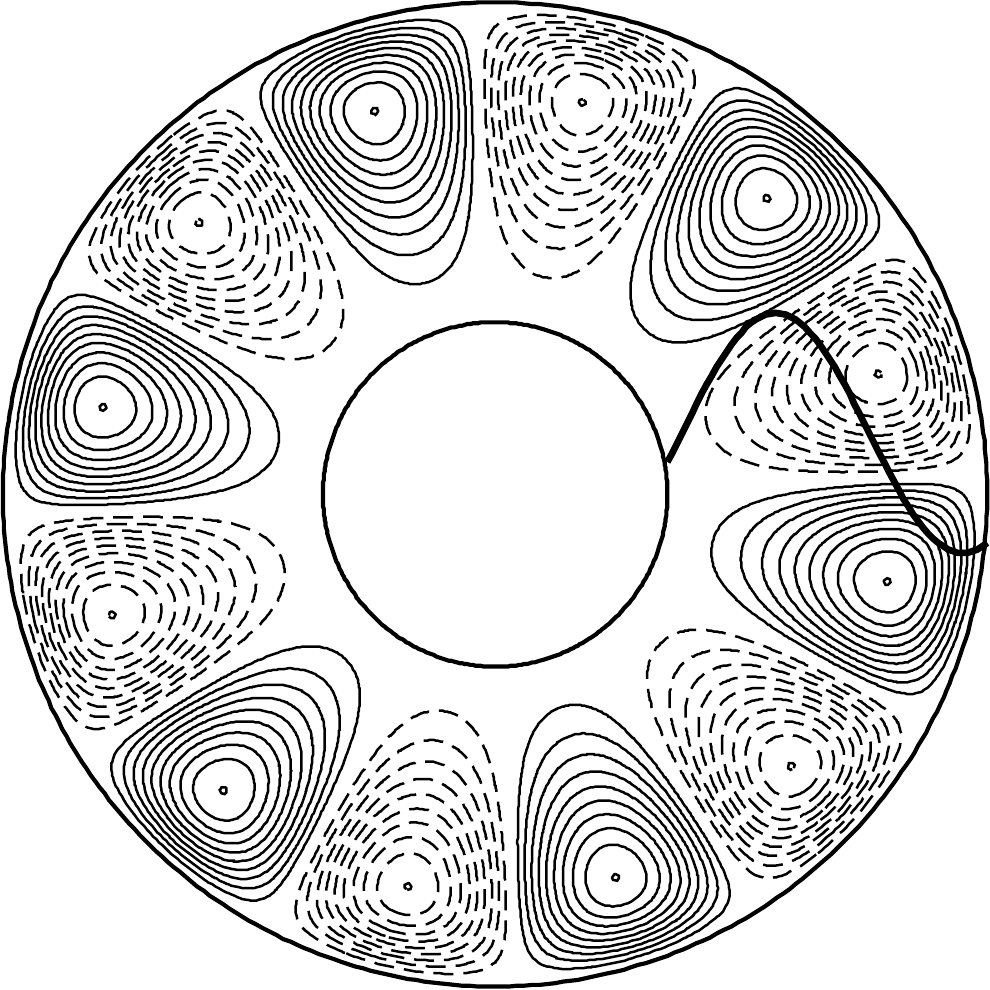}}
\end{minipage}
\\[2mm]
\begin{minipage}{0.49\linewidth}
\centerline{$\Psi$, $\lundquist=10^2$, slow mode}
\centerline{\includegraphics[clip=true,width=\mylw]{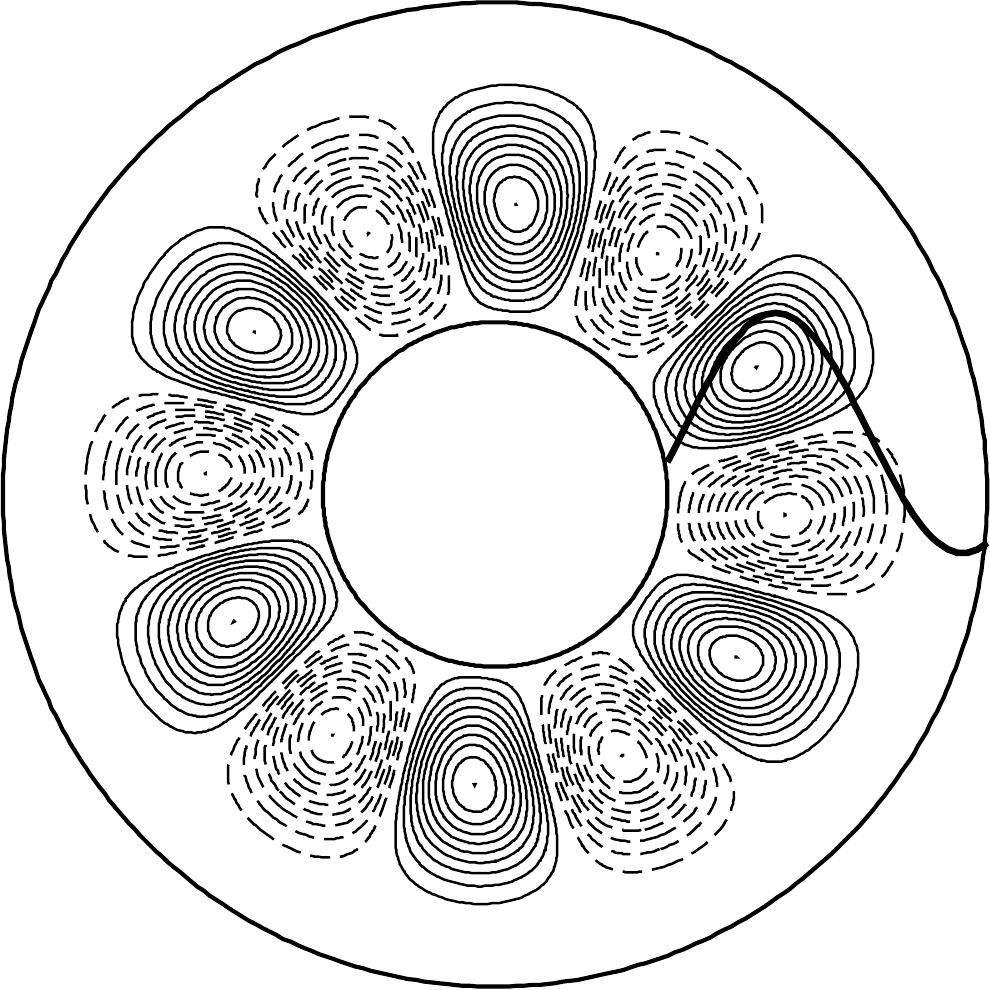}}
\end{minipage}
\hfill 
\begin{minipage}{0.49\linewidth}
\centerline{$\Psi$, $\lundquist=10^2$, fast mode}
\centerline{\includegraphics[clip=true,width=\mylw]{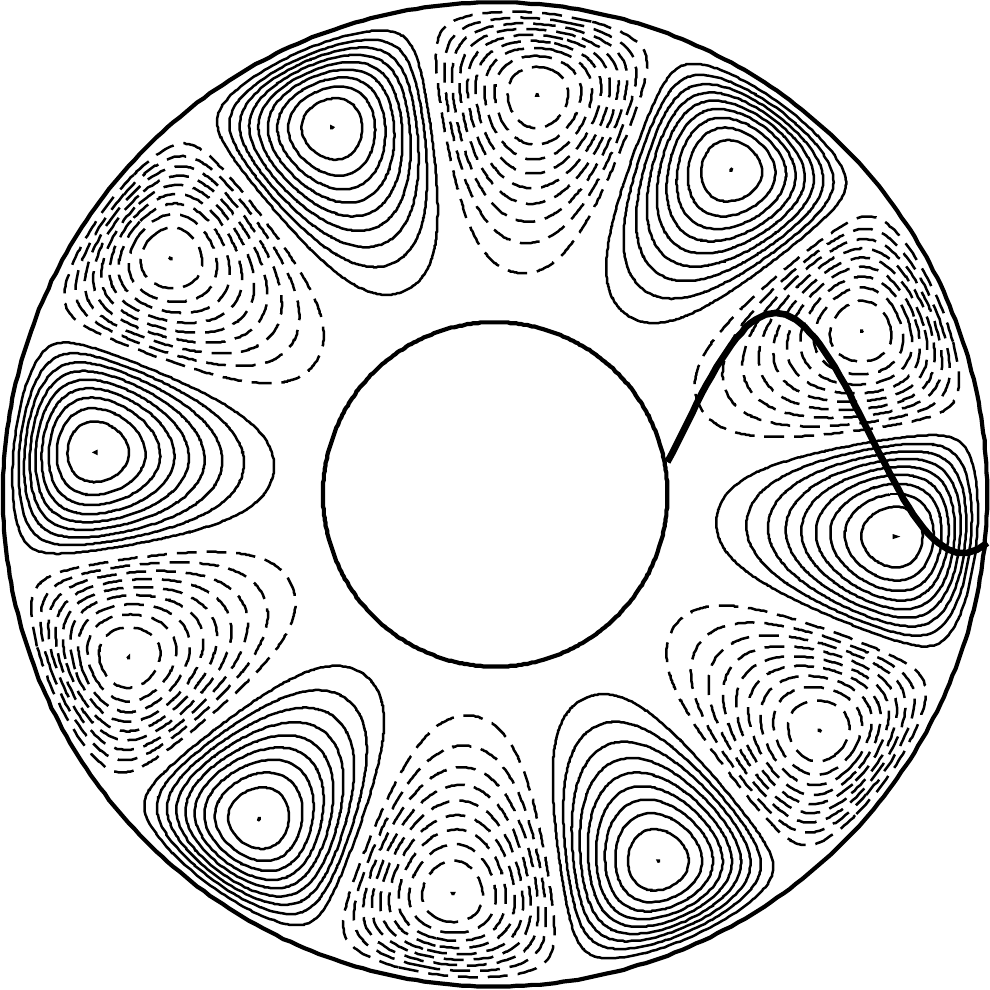}}
\end{minipage}

\caption{Contours of the velocity streamfuction $\Psi$ in the equatorial plane as seen 
from above. Shape of the fundamental eigenfunction of azimuthal wavenumber $m=6$ for the
slow mode (left panels) and the fast mode (right panels) in spherical shell geometry in the presence of the general background field. The first row corresponds to no magnetic dissipation whereas the second and third rows are for $\lundquist=10^{4}$ and $\lundquist=10^2$ respectively. All figures are $Le=10^{-4}$. All streamfunctions are 
normalized, and the plots show ten equally-spaced levels in positive (solid contours) and negative (dashed contours). The thick black line shows the background magnetic field profile. }
\label{MRmagDissip1}
\end{minipage}
\end{center}
\end{figure}

\begin{figure}
\begin{center}
\begin{minipage}{.9\linewidth}
\begin{minipage}{0.49\linewidth}
\centerline{$A$, $\lundquist=+\infty$, slow mode}
\centerline{\includegraphics[clip=true,width=\mylw]{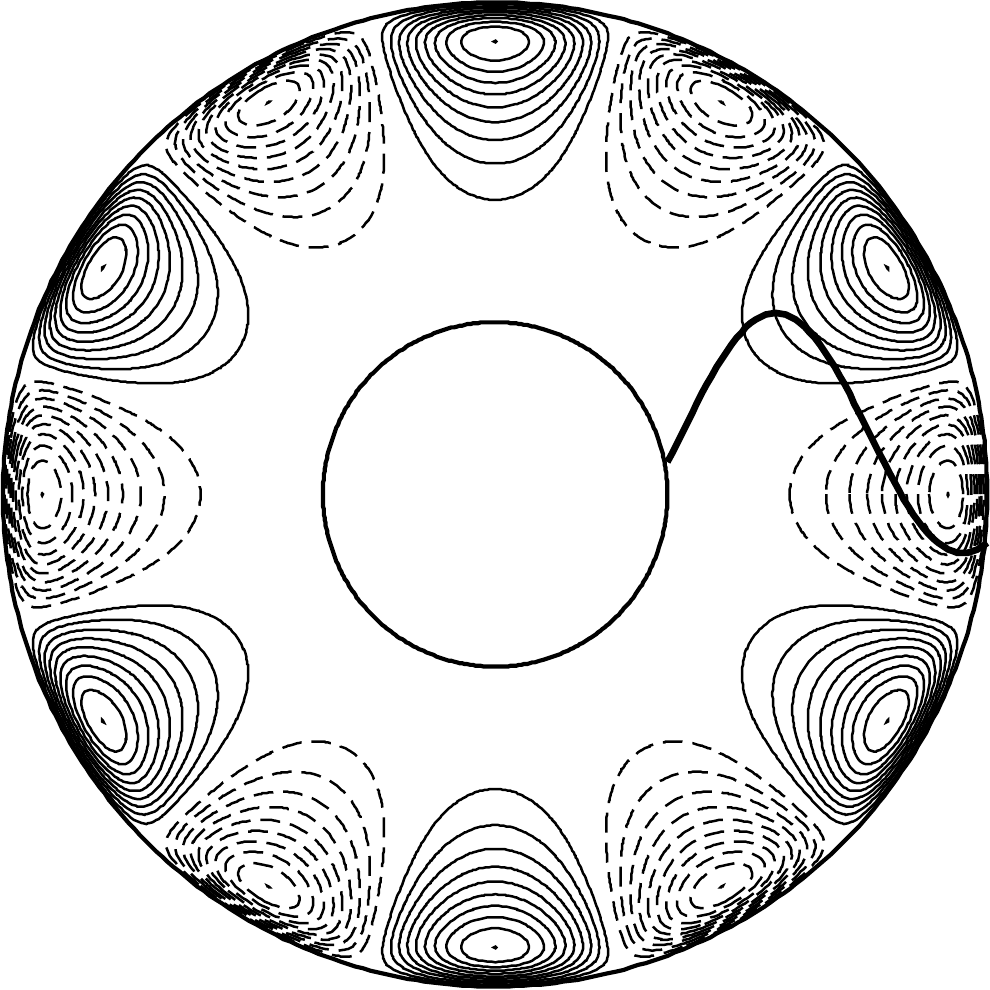}}
\end{minipage}
\hfill 
\begin{minipage}{0.49\linewidth}
\centerline{$A$, $\lundquist=+\infty$, fast mode}
\centerline{\includegraphics[clip=true,width=\mylw]{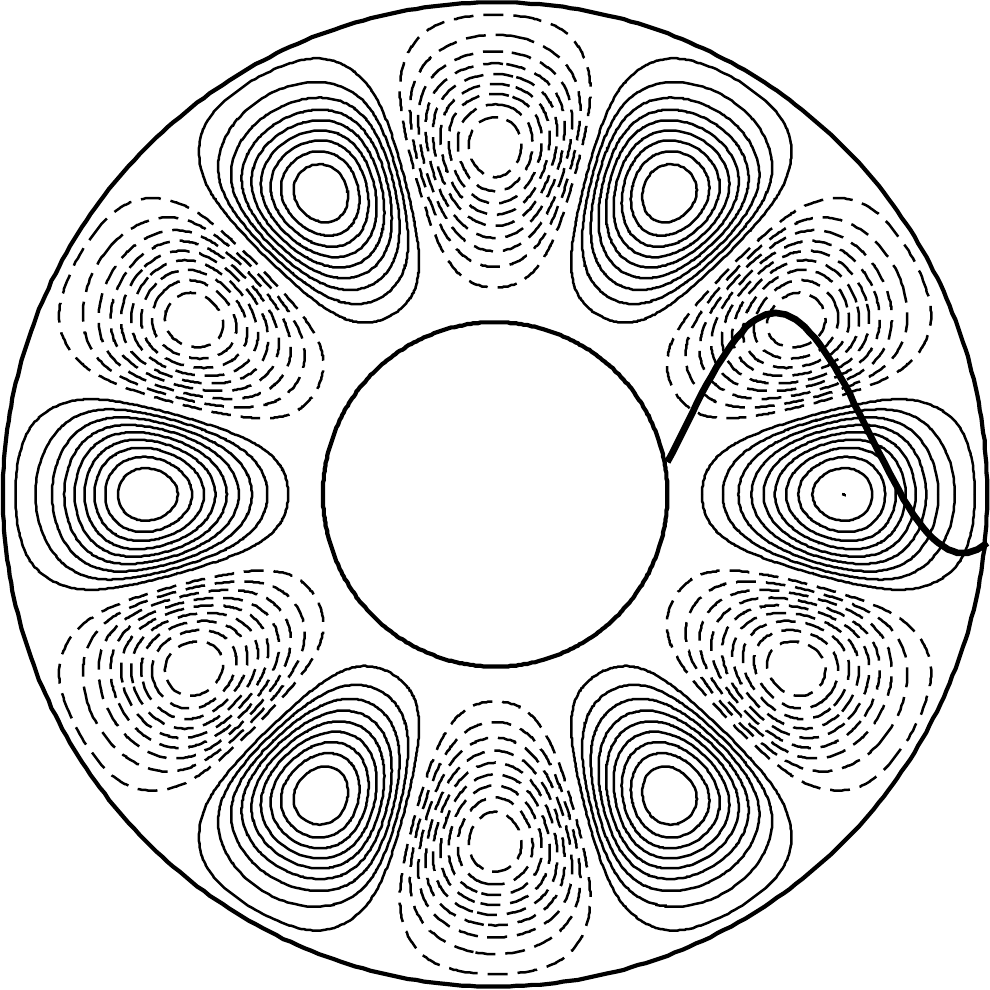}}
\end{minipage}
\\[2mm]
\begin{minipage}{0.49\linewidth}
\centerline{$A$, $\lundquist=10^4$, slow mode}
\centerline{\includegraphics[clip=true,width=\mylw]{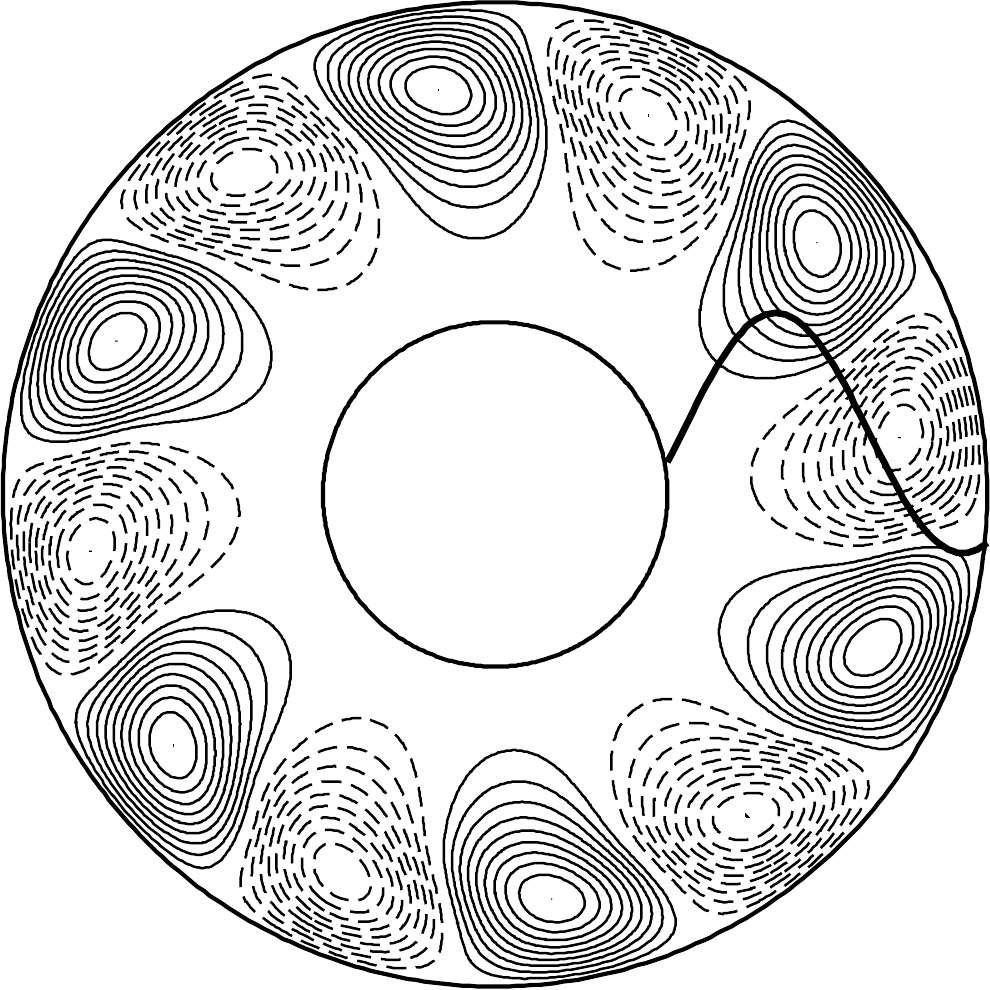}}
\end{minipage}
\begin{minipage}{0.49\linewidth}
\centerline{$A$, $\lundquist=10^4$, fast mode}
\centerline{\includegraphics[clip=true,width=\mylw]{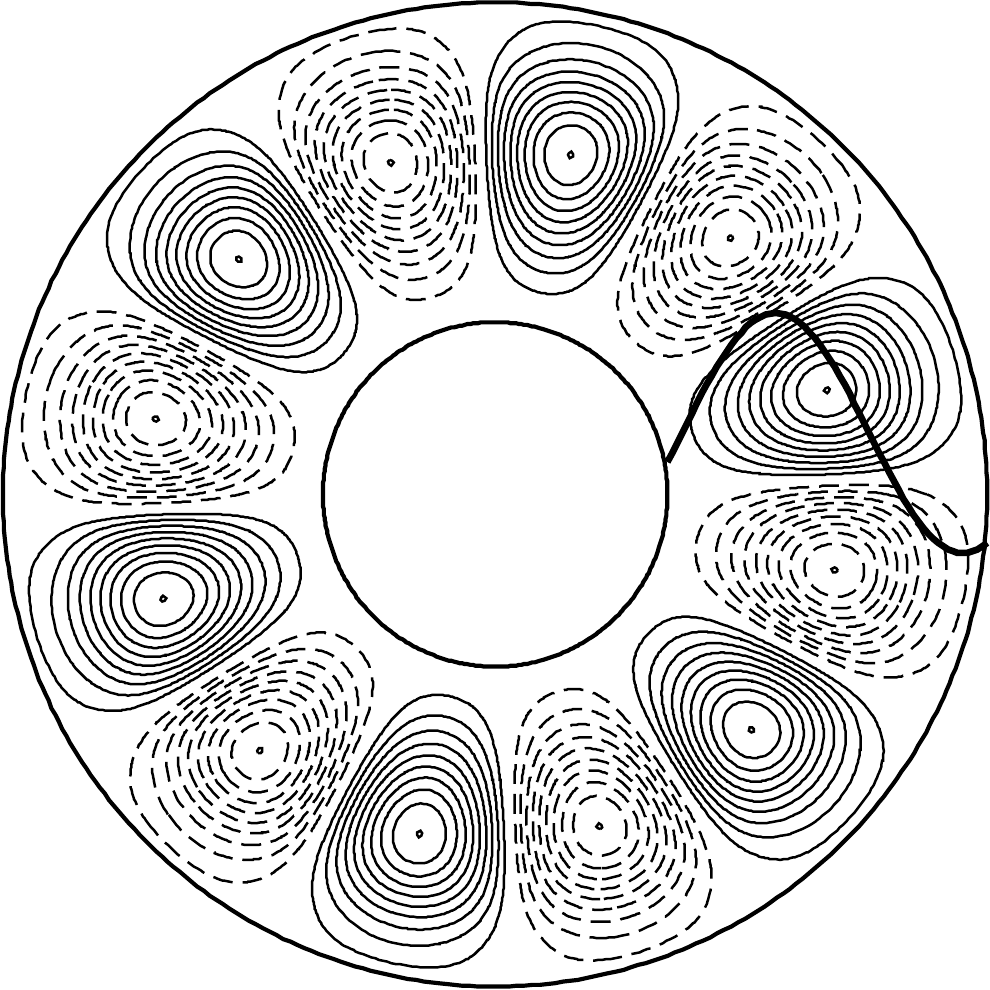}}
\end{minipage}
\\[2mm]
\begin{minipage}{0.49\linewidth}
\centerline{$A$, $\lundquist=10^2$, slow mode}
\centerline{\includegraphics[clip=true,width=\mylw]{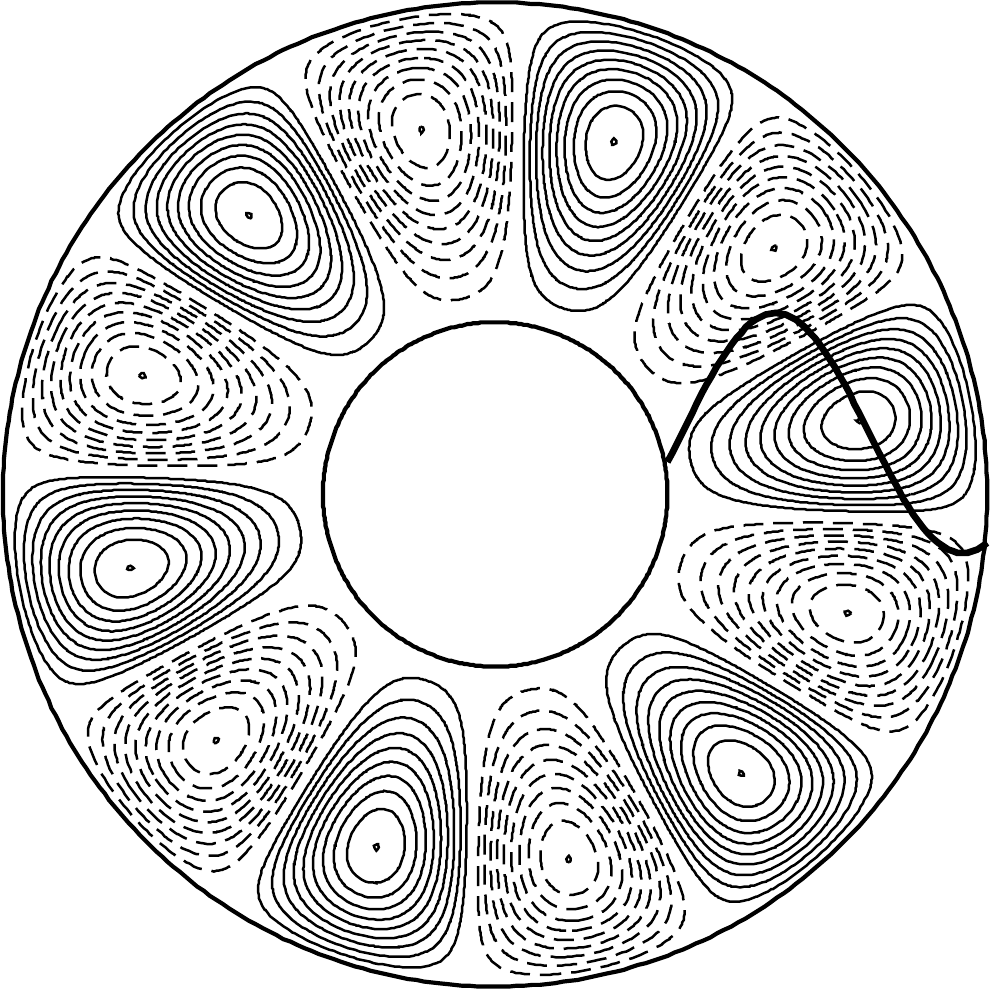}}
\end{minipage}
\hfill 
\begin{minipage}{0.49\linewidth}
\centerline{$A$, $\lundquist=10^2$, fast mode}
\centerline{\includegraphics[clip=true,width=\mylw]{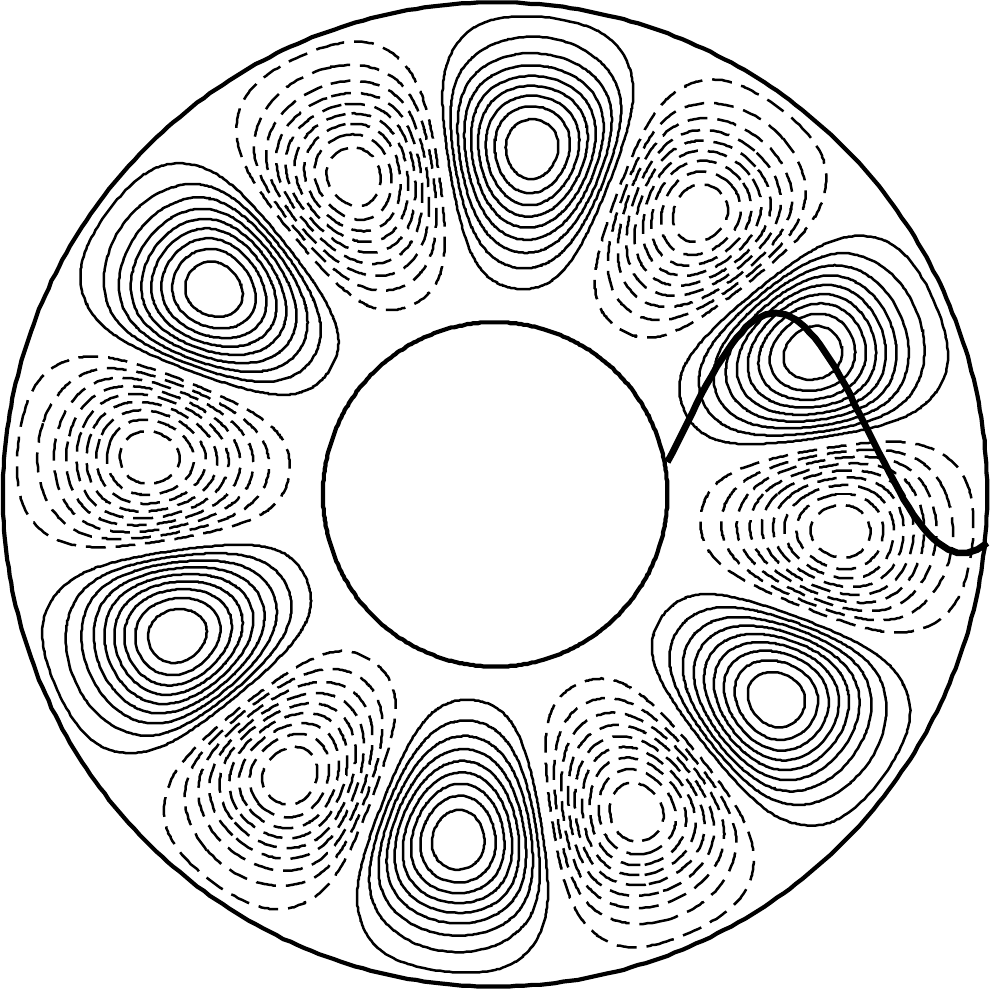}}
\end{minipage}
\caption{Contours of the magnetic streamfuction $A$ in the equatorial plane as seen from above. Shape of the fundamental eigenfunction of azimuthal wavenumber $m=6$ for the slow mode (left panels) and the fast mode (right panels) in spherical shell geometry in the presence of the general background field. The first row corresponds to no magnetic dissipation whereas the second and third rows are for $\lundquist=10^{4}$ and $\lundquist=10^2$ respectively. All figures are for $\lehnert=10^{-4}$. All streamfunctions are normalized, and the plots show ten equally-spaced levels in positive (solid contours) and negative (dashed contours). The thick black line shows the background magnetic field profile.  }
\label{MRmagDissip1_field}
\end{minipage}
\end{center}
\end{figure}

\begin{figure}
\begin{minipage}{0.49\linewidth}
\centerline{\includegraphics[clip=true,width=0.95\linewidth]{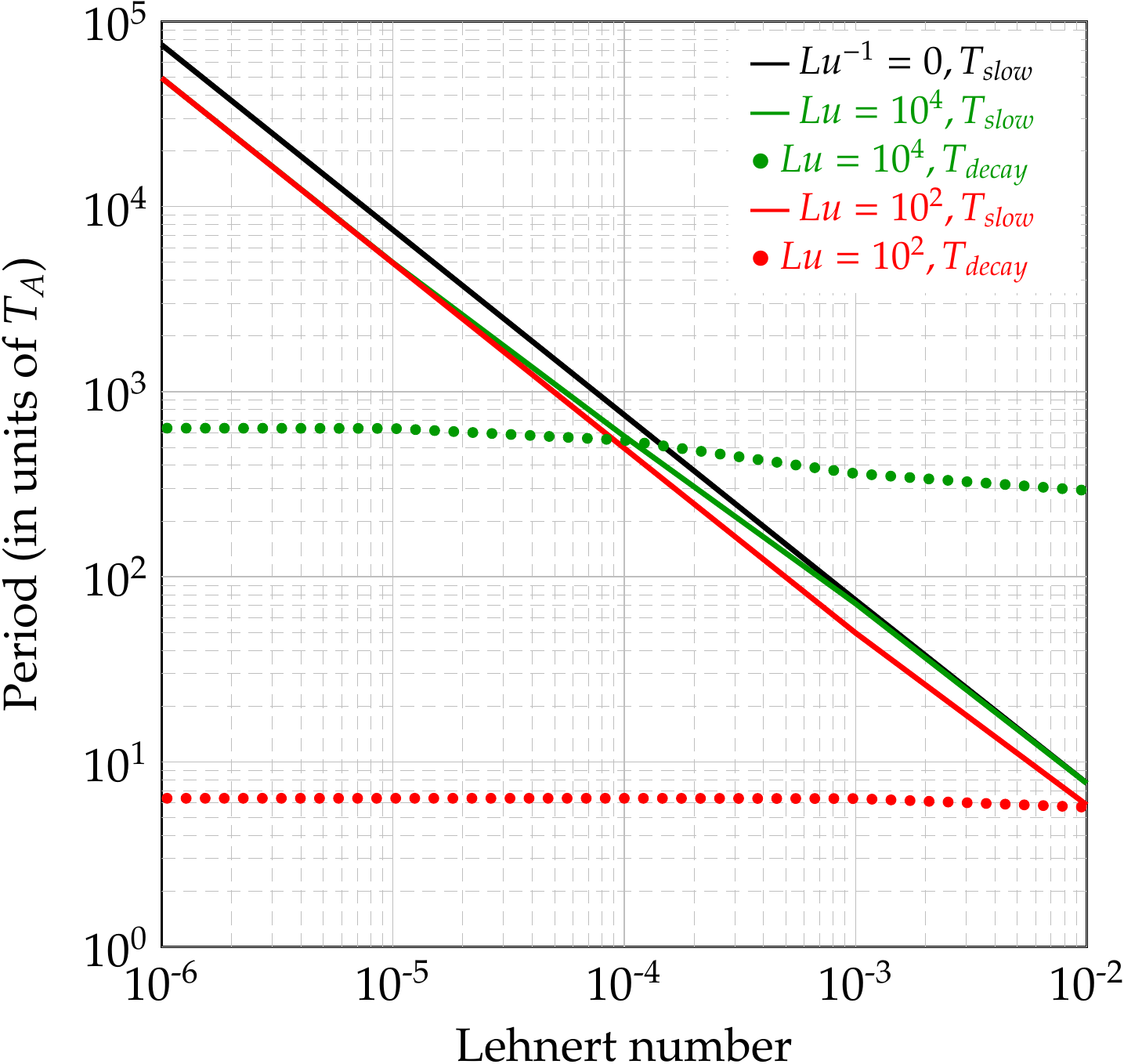}}
\end{minipage}
\begin{minipage}{0.49\linewidth}
\centerline{\includegraphics[clip=true,width=0.95\linewidth]{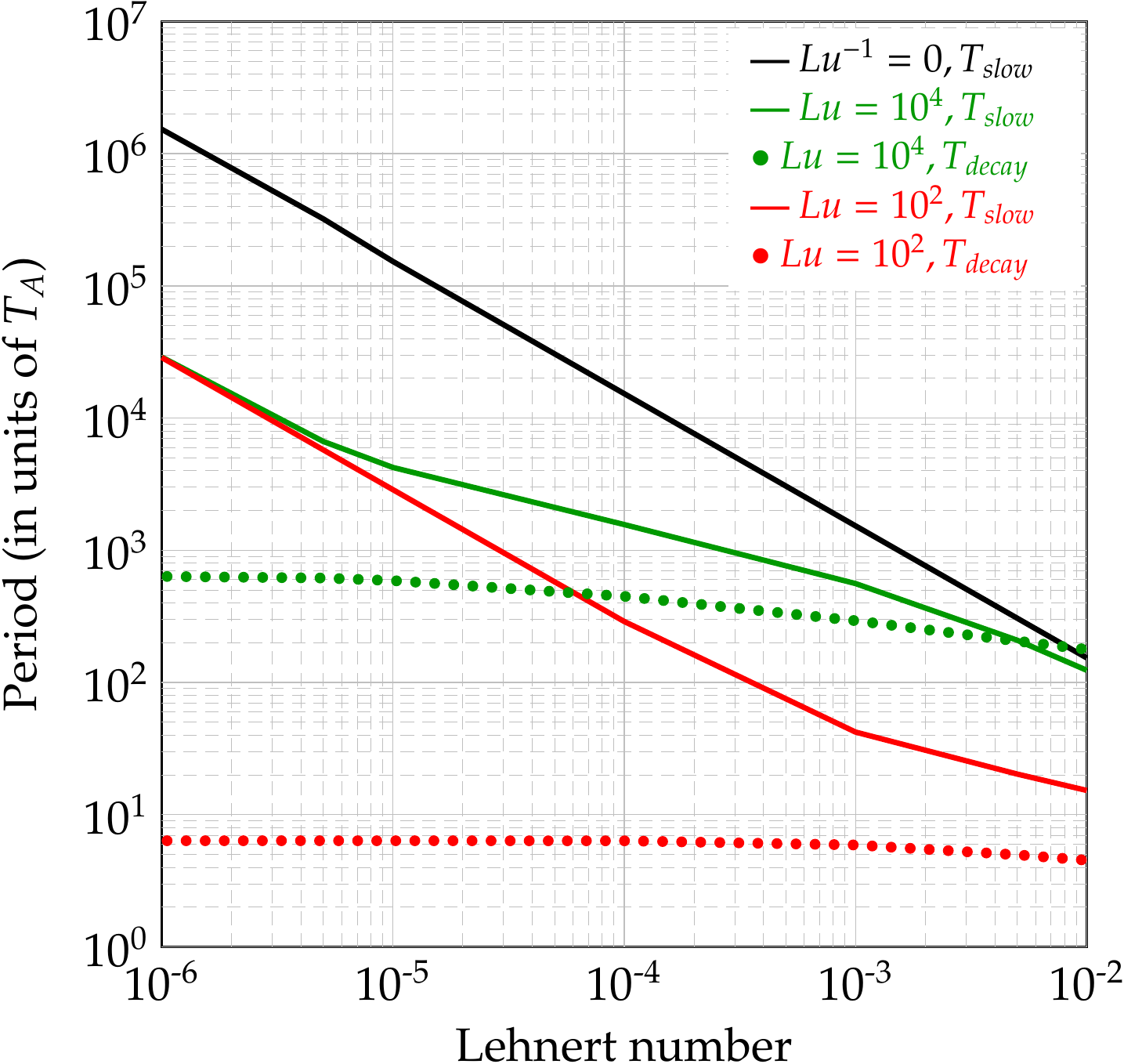}}
\end{minipage}
\caption{Oscillation (pseudo) periods and decay times for the modes  as a function of $\lehnert$, for the fundamental mode in radius with azimuthal wavenumber $m=6$. Black solid curves for modes without dissipation $\lundquist=+\infty$, Green curves for weak magnetic dissipation $\lundquist=10^{4}$ (solid: period, bullets: decay time), blue curves for strong magnetic dissipation $\lundquist=10^{2}$ (solid: period, bullets: decay time). The background field is the Malkus field (left panel) and the more general zonal field (right panel) respectively.}
\label{MRmagDissip2}
\end{figure}

Figure~\ref{MRmagDissip1} presents contours of the velocity stream function in the equatorial plane for the slow (left panels) and fast (right panels) fundamental hydromagnetic modes for azimuthal wavenumber $m=6$ with various amount of magnetic dissipation present.  The upper panels show the diffusionless case ($\lundquist =+\infty$), while the middle and lower panels involve $\lundquist=10^{4}$ and $\lundquist=10^{2}$ respectively.
The basic structure of the fast modes is essentially unchanged by the introduction of magnetic dissipation although the velocity streamfunctions do show a small phase shift that depends on the amount of dissipation but is independent of the background magnetic field.
The structure of the slow mode, on the other hand, is strongly influenced by the dissipation. At $\lundquist =10^{4}$, the velocity field spirals as a function of cylindrical radius, while if dissipation is increased further up to $\lundquist = 10^{2}$ it becomes straight again but now shifted toward the inner core suggesting another mechanism then controls the mode structure. 

Figure \ref{MRmagDissip1_field} presents similar plots of the magnetic streamfunction.  Note the magnetic and velocity stream functions are now no longer identical, as was the case in the study of \citet{m1967} and for the analytic solutions we presented above, instead the velocity stream functions are now noticeably more compressed towards the core-mantle boundary than the magnetic stream functions. Considering the fast modes, the magnetic and velocity streamfunctions are found to be in anti-phase.  Considering the slow modes, the magnetic and velocity streamfunctions are in phase for the diffusionless and weak diffusion cases, but exactly out-of-phase when the magnetic diffusion is strong (Lu=100). 
In both Figure~\ref{MRmagDissip1} and Figure \ref{MRmagDissip1_field} the background magnetic field is plotted as the thick black line.  It seems to dictate where spiralling of the modes (both in the velocity field and the magnetic field) takes place in the intermediate regime of weak magnetic diffusion. In particular, with the general zonal background field investigated here, the spiraling pattern is narrow and located where the gradient of the background field is large. With the Malkus background magnetic field the spirals are broader and fill the whole domain (not shown).

Figure~\ref{MRmagDissip2} documents the influence of magnetic dissipation ($\lundquist=10^{2},10^{4}$) or no dissipation on mode periods for the case of the Malkus (left panel) and more general (right panel) background zonal magnetic fields.  The crucial change compared to the diffusionless case is that the eigenfrequency becomes complex.   The real part of the eigenfrequency then defines the pseudo-period or oscillations period of the mode, for example,
\begin{equation}
T_{slow}=\frac{2\pi}{\mathrm{Re} \left\{\omega \right\}}
\end{equation}
while the imaginary part of the eigenfrequency corresponds to its decay time
\begin{equation}
T_{decay}=\frac{2\pi}{\mathrm{Im} \left\{\omega \right\}}.
\end{equation}
In both panels, there are regions of parameter space where the decay timescale $T_{decay}$ is comparable to, or shorter than, the pseudo-period $T_{slow}$. 
In such cases, the mode will have no time to oscillate before it is dissipated.

\section{Discussion}
\label{sec:discussion}

A fundamental assumption in this study is that the axially invariant QG model is an adequate description of transient flow disturbances on time scales slower than the rotation time scale.  Detailed comparisons with slow (three-dimensional) inertial waves in a sphere (see \ref{app:zhang}) suggest that this is indeed the case in the hydrodynamic scenario, provided that one considers large scale (low $m$) motions that are slow enough for inertia to be negligible.  In the hydromagnetic case, a further requirement is that the Lorentz force should remain much weaker than the the Coriolis force.  At the very low Lehnert numbers of $10^{-3} - 10^{-6}$ relevant for planetary cores this is the case, even if the background magnetic field is fully three dimensional \citep{g2011}. A more troublesome assumption is that the magnetic field perturbation is also axially invariant.  This is likely reasonable in the diffusionless limit, and probably also when magnetic diffusion is weak, but when magnetic diffusion is strong it will certainly   break down.  In the quasi-free decay regime (which is likely of limited dynamical relevance because the decay time scale of these modes is shorter than their oscillation time scale) full three-dimensional models such as described by \cite{s2012} should be preferred to a QG approach.

Interestingly, the QG model implemented here is not restricted to a spherical shell, but can be used to study any axisymmetric container.  This allows comparisons to be made with earlier annulus type models with constant boundary slope.  The spherical shell geometry appropriate for planetary cores, with the boundary slope increasing as a function of the cylindrical radius, causes eigenfunctions to become more compressed towards the outer boundary at large cylindrical radius, particularly when higher radial overtones are considered.  If the background magnetic field also varies with cylindrical radius, this provides an additional modulation of the radial structure of the modes.  As illustrated in Figure \ref{allfields2}, the mode amplitude tends to be enhanced where the magnetic field is weaker.  Since in planetary cores the toroidal field is expected to decay towards the outer boundary (since silicate mantles are relatively poor electrical conductors) this will also tend to push the mode minima and maxima out to large cylindrical radius.  A prediction of our QG model of  hydromagnetic modes in a spherical shell is therefore that the free oscillation modes are more likely to be observed at mid to low latitudes.


Perhaps the most important finding of this study is that magnetic dissipation will strongly influence the properties of slow magnetic modes in planetary cores.  In our numerical experiments a transition from diffusionless slow modes to slow modes dominated by diffusion (referred to as quasi free decay (QFD) modes by \cite{s2012}) occurs as the Lehnert number becomes smaller than the inverse Lundquist number.  For lower Lehnert numbers, the decay time of the slow modes is much shorter than their oscillation time, so they will rapidly decay and are precluded from playing an important role in the core dynamics or producing observable secular variation.  The transition from diffusionless to QFD slow magnetic modes happens at higher Lehnert number when the background magnetic field is more complex in the radial direction.  Since $\lehnert \cdot \lundquist=\elsasser$ is the traditional Elsasser number, another way to state this is that for $\elsasser \ll 1$, free slow magnetic modes are expected to rapidly decay.  This conclusion was essentially arrived at previously by \cite{s2012} in his study of a small number of 3D modes, here we confirm this finding for our more extensive catalogue of QG modes.

\begin{table}{}
\centering
\renewcommand{\arraystretch}{1.1} 
\setlength{\tabcolsep}{0.1cm}
\begin{tabular}{|c||c|c|c|c|c|c|c|c|c|c|} \hline
&$m$ & \multicolumn{3}{c|}{$m=3$} &   \multicolumn{3}{c|}{$m=6$}  &   \multicolumn{3}{c|}{$m=12$} \\
&radial mode & 1 & 2 & 4  & 1 & 2 & 4 & 1 & 2 & 4 \\\hline 
\multirow{2}{*}{Malkus}&$T_{slow}$ (yrs) & $10300$& $3100$& $1244$ &$3080$& $1310$& $529$& $833$& $380$ & $171$\\
&$T_{decay}$ (yrs) &  $41750$& $13900$&$4013$ &$15500$& $6800$& $2707$&$4800$& $2400$& $1212$ \\\hline 
\multirow{2}{*}{General}&$T_{slow}$ (yrs)& $43800$& $7600$& $1840$ &$24000$&$5000$& $1260$&$11100$&$2900$ & $794$\\ 
&$T_{decay}$ (yrs) & $24000$& $5300$& $1478$ &$12600$&$3400$& $969$&$5700$&$2000$ & $627$\\ \hline
\end{tabular}
\caption{Pseudo-period $T_{slow}=\frac{2\pi}{\mathrm{Re}(\omega)}$ and decay time $T_{decay}=\frac{2\pi}{\mathrm{Im}(\omega)}$ in years for Earth's core (taking $\lehnert=10^{-4}$ and $\lundquist=10^5$, a value representative of the range estimated for Earth's core $\left[59,000-186,000\right]$ assuming the core interior  field strength of $3$~mT proposed by \citet{g2010}).  Results for slow QG modes with azimuthal wave numbers $m=3, 6, 12$ and three radial modes (the 1st, 2nd and 4th modes) are presented.  Results are shown for both the Malkus general field and the more realistic general zonal field.  To obtain these times in years, the non-dimensional results from the numerical model were rescaled using $1 T_A=4.51$~years, as is appropriate for $\lehnert=10^{-4}$.}
\label{tabledissip}
\end{table} 

\begin{figure}
\begin{minipage}{0.59\linewidth}
\centerline{\includegraphics[width=.95\linewidth]{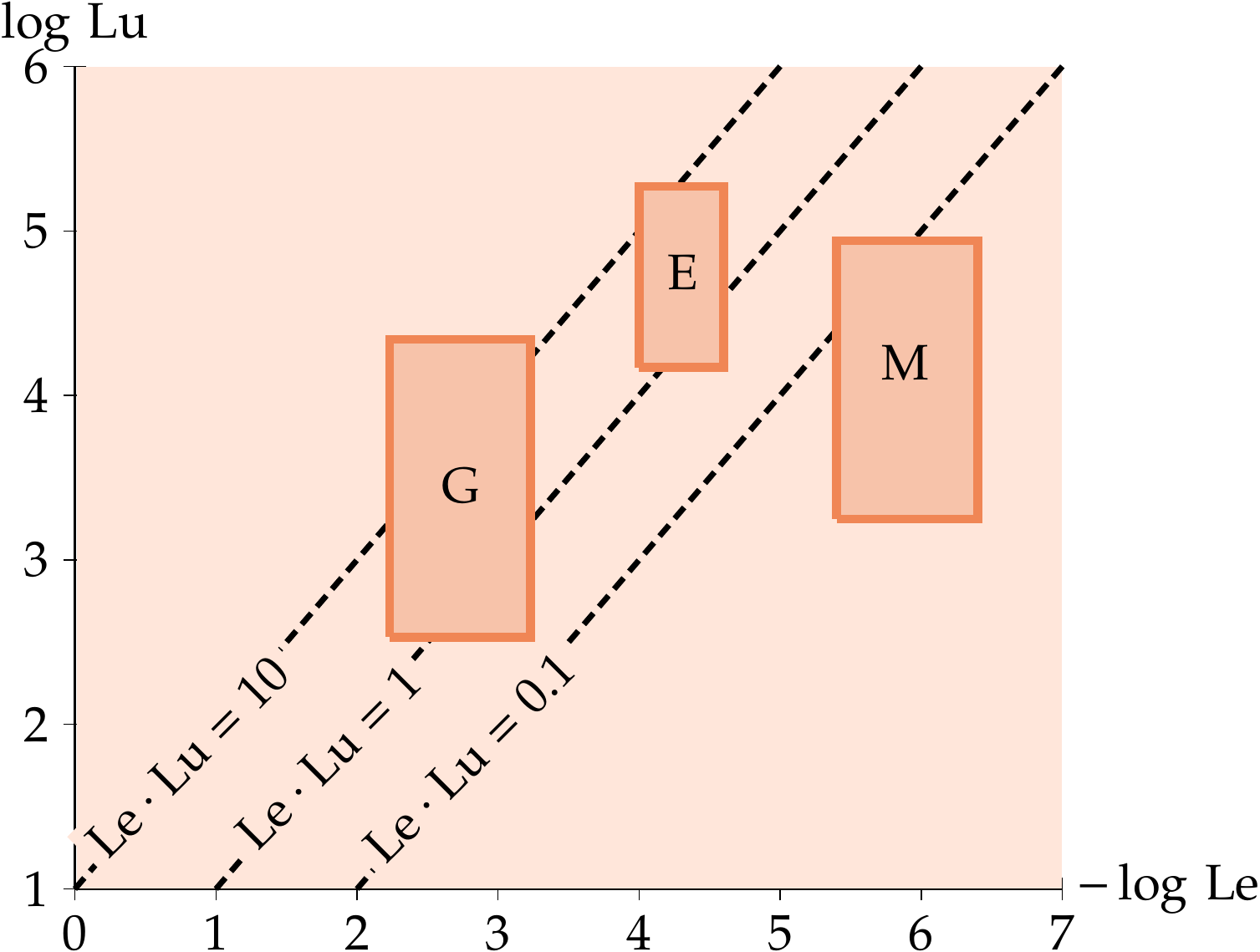}}
\caption{Approximate location of Earth (E), Ganymede (G) and Mercury (M) in the $\lehnert-\lundquist$ plane (logarithmic scales). The estimates come from the data presented in table~\ref{tab:Lehnert}.}
\label{fig:LeLu}
\end{minipage}
\end{figure}

For the Earth's core we estimate the Lehnert number lies in the range $\lehnert= \left[2.5\cdot 10^{-5} -10^{-4} \right]$ while the Lundquist number is estimated to lie in a range $\left[14, 750 - 186,000\right]$ so the Elsasser number is in the range $\left[0.4 - 18.6\right]$ (see Table~\ref{tab:Lehnert}).   It is therefore possible that a transient slow hydromagnetic mode might be observed for a few periods before it decays, especially if the strong field proposed by \citet{g2010} and the most recent values for core conductivities \citep{dekoker2012,pozzo2012}, which lead to the upper estimates of $\lehnert$ and $\lundquist$, prove to be correct.  However, this will also depend on the complexity of the background magnetic field.  This point is demonstrated in Table \ref{tabledissip} where the pseudo-periods and decay times in years, rescaled using parameters appropriate for the Earth's core, are presented for the Malkus background field and the general zonal background field.  For the Malkus background field, we find decay times longer than the oscillation times while for the general zonal background field the opposite is true.  The decay time approaches the oscillation time for higher overtones, so these may still be of observational relevance. It is the low Lehnert number in planetary cores that ensures the slow modes have a magnetic energy much larger than their kinetic energy - see \eqref{energy_ratio}. This underlies the important role played by magnetic diffusion for the slow modes.  The opposite is true for the fast modes which have kinetic energy much larger than their magnetic energy.  An additional source of dissipation that may be important in planetary cores, but which has for simplicity been neglected here, is boundary coupling.  For example,  if the lowermost mantle involves a layer with high electrical conductivity then the magnetic field will penetrate into it and the differential motion between the core fluid and the mantle will shear the magnetic field dissipating energy. \cite{dumberry2008} investigate this source of damping in a study of torsional oscillations.  Consideration of this mechanism would require a modification of our simple boundary conditions, and may further shorten the decay time scale for the slow magnetic modes. 

It is noticeable that the periods of the large-scale slow magnetic modes obtained using the general zonal background field (see table \ref{tabledissip}) are rather longer than the time scales of order 300 years normally associated with westward drifting flux patches at Earth's core surface.  Furthermore, in his pioneering study, \cite{h1966} highlighted somewhat shorter periods for the slow magnetic modes.  We find that the periods of the slow magnetic modes are very sensitive to both the shape and the strength of the background magnetic field, for example, even in the simplest case the dimensional period of the slow mode scales as $B_0^2$ (in (\ref{eq:thTs}), units of $T_A$ are used), so a change in the field strength by only a factor of 3 would result in an order of magnitude change in the mode period.  For our specific choice of background field structure and strength (see table \ref{tabledissip}) some westward bulk advection would be necessary in addition to the QG modes in order to explain the observed rate of westward drift.  More accurate conclusions must await an improved knowledge of the background magnetic field in Earth's core.  It has not escaped our attention that this sensitivity of mode structure to the background magnetic field may provide a means of probing (inverting) for magnetic field structure within Earth's core. 

Figure~\ref{fig:LeLu} presents the location of Earth, Ganymede and Mercury in a $\lehnert-\lundquist$ plane, about a $\elsasser = 1$ straight line.
Our results imply that  free oscillation, slow, magnetic modes will be rapidly damped by magnetic dissipation in the core of Mercury, while in Earth and possibly Ganymede QG modes modified by diffusion may persist for a few periods before decaying. The fast modes, on the other hand are only weakly influenced by magnetic dissipation and are expected to be present in all the planetary cores considered.  If hydromagnetic modes were continuously or intermittently forced in the cores of Earth or Ganymede they could play an important role in transient dynamics, particularly if the Elsasser number for these bodies turns out to be at the larger end of current estimates.  It will nonetheless be a major observational challenge to conclusively observe these hydromagnetic modes.  One requires globally distributed magnetic observations for a time scale comparable to their period.  For the slow magnetic modes, this is hard to envisage except perhaps for the Earth.  But for the fast modes observational identification using magnetic measurements from orbital satellites may be mor feasible, provided mantle conductivities of the planets are not so large as to filter out the mode signatures, and provided the external (non-core) magnetic field fluctuations can be efficiently separated at the periods of the modes.

In an effort to focus on the intrinsic properties of QG modes, no excitation source was included in our model.  The logic of first establishing the properties of the free oscillations and then using them within a Green's function framework, with a particular excitation source, is well known from normal mode seismology and more recently from study of torsional oscillations in Earth's core \citep{buffett2009}.  A similar development could be envisaged for the non-axisymmetric hydromagnetic QG modes discussed here, although their non-normal nature should be borne in mind.  What are the possible excitation sources?  Most obviously there is thermochemical core convection.  The time scales of several hundred years for convective overturn in the Earth's core are appropriate for exciting axially invariant motions and QG hydromagnetic modes.  Other possibilities include shear flow instability or instability of magnetic field itself.  A final intriguing possibility is topographic forcing along the lines originally proposed by \cite{h1966}. The time scale and spatial structure of the forcing mechanisms (e.g.  fluctuations in columnar convection outside the tangent cylinder, shear instability near the tangent cylinder, or magnetic field instability at low latitudes close to the outer boundary where the shear in the toroidal field may be large) will determine which modes are preferentially excited and possibly maintained against diffusion long enough to be observed.

Linear models of waves and modes, such as that presented here, have proven important in obtaining physical insight in other geophysical and planetary scale fluid systems, including  Earth's atmosphere and oceans.  But to explain geophysical observations in detail it is usually necessary to go beyond such simple linear models, and to consider not only excitation sources but also coupling, feedback and saturation between the mean state and the disturbances.  The linear model presented here is not suitable for such studies, these  require fully nonlinear QG models of core dynamics that are currently under development.  Indeed the long periods obtained for the slow magnetic modes in this study lead us to expect that the background magnetic field will evolve on a time scale similar to that of the slow modes.  Nonlinear models are thus essential for an assessment of the physical relevance of the slow magnetic modes.
Such nonlinear models will more generally enable us to answer important questions such as whether transient hydromagnetic modes riding apon large-scale gyres (generated by convection in planetary cores), could have an observable imprint on the magnetic field variations of these bodies. 

\section*{Acknowledgements}
We thank Mathieu Dumberry and an anonymous reviewer for constructive and insightful comments that helped to improve the manuscript.  
This study has benefited from the support of the International Space Science Institute (ISSI), through the work of teams 176 and 241. E.C. was supported by the ETH Z\"urich Postdoctoral fellowship Progam during the early stages of the study. C.F. was kindly supported by IPGP during a visit to Paris in June 2010.  A.F. acknowledges support from the Centre National d'Etudes Spatiales. The contribution of A.F. is IPGP contribution 3466.

\bibliographystyle{elsarticle-harv}
\bibliography{./CF_QG_refs_new.bib,./hdrplus,./EC_QG_refs.bib}

\begin{appendix}

\section{Comparison between quasi-geostrophic modes and slow inertial modes in a sphere}
\label{app:zhang}
In this appendix, we present a comparison between free oscillations 
predicted by our QG model with a spherical outer boundary, and previously derived analytical solutions for slow inertial modes in a full sphere \citep{z2001}.  The later solutions involve no a priori assumption of quasi-geostrophy (i.e. no axial averaging).   This comparison is an interesting test of our QG model, providing a quantitative benchmark against known solutions, and allow us to ascertain the conditions under which the QG approximation breaks down.

It has long been recognized that Rossby modes in a closed container are no more than a special class of inertial modes, in which ${\partial {\bf u} / \partial t}$ is much smaller than the pressure and Coriolis forces, so that related motions are very close to geostrophic \citep{lh1964,g1968,b2005}.  Although we present a comparison here of purely hydrodynamic modes, it is interesting to note that in the case of an imposed Malkus magnetic field \eqref{eq:malkusf}, then a very similar comparison can also be made for hydromagnetic modes \--- see \cite{m1967,z2003}.  

Since the comparison in this section relies on the inertial mode solutions of \citet{z2001}, we briefly remind the reader of their salient characteristics.  

The inertial modes in a sphere are the solutions to
\begin{equation}
\partial_t \mathbf{u}   + 2 \zv \times \mathbf{u} = - \boldsymbol{\nabla} p   
\label{IW_prob}
\end{equation}
and 
\begin{equation}
\boldsymbol{\nabla} \cdot \mathbf{u} = 0, 
\end{equation}
subject to the impenetrable boundary condition $u_r=0$ on the the spherical surface at $r=r_o$.  To facilitate comparisons with \citet{z2001}, the rotation time scale $\Omega^{-1}$ has been used here to non-dimensionalize.   Expanding these vector equations into cylindrical co-ordinates, considering solutions of the form
\begin{equation}
\mathbf{u}(s,\varphi,z)=\left[ U_s(s,z), U_\varphi(s,z), U_z(s,z)\right] e^{\mathrm{i}(m \varphi + 2 \sigma t)},
\label{IW_exp}
\end{equation}

and substituting between the components to eliminate pressure, \cite{z2001} show how the half frequencies $\sigma$ for equatorial symmetric solutions i.e. for which $(U_s,U_\varphi,U_z)(\cdot,z)=(U_s,U_\varphi, -U_z)(\cdot,-z)$ (the symmetry compatible with the QG model) can be found from the roots of the equation

\begin{equation}
\sum^N_{j=0} (-1)^{j} \frac{[2(2N+m-j)]!}{j!(2N+m-j)![2(N-j)]!}\left[ (m +2N-2j) - \frac{2(N-j)}{\sigma_{Nmn}}\right] \sigma_{Nmn}^{2(N-j)}=0.
\label{Zhang_ES_roots}
\end{equation}

Different solutions are obtained for $N=0,1,2,\dots$ and $n$ indexes the possible modes for a given choice of $N$ and $m$.  \citet{z2001} describe how in the triplet $\left\{Nmn\right\}$, $m$ denotes the azimuthal wavenumber, $N$ defines the degree of possible complexity in the axial ($z$) direction and $n$ is related to the cylindrical radial structure. 

Considering the special case of slow inertial modes for which 
\begin{equation} \sigma_{si} = \sigma_{Nmn} \ll \mathcal{O} (m/(N+m),\end{equation}
that are close to geostrophy and so possess relatively simple structures in the axial $z$ direction, then an analytic approximation to (\ref{Zhang_ES_roots}) is available \citep{z2001,b2005}
\begin{equation}
\omega_{si} \simeq -{2 \over m+2} \left[\sqrt{1+ \frac{m(m+2)}{N(2N+2m+1)}} -1\right],
\end{equation}
where $\omega_{si}=2 \sigma_{Nmn}$ and this approximation is exact for the case $N=1$.  All such slow modes propagate eastwards and are none other than the well known Rossby modes in a full sphere\citep{b2005}.

\begin{figure}
\begin{minipage}{.49\linewidth}
\centerline{\includegraphics[width=0.95\linewidth]{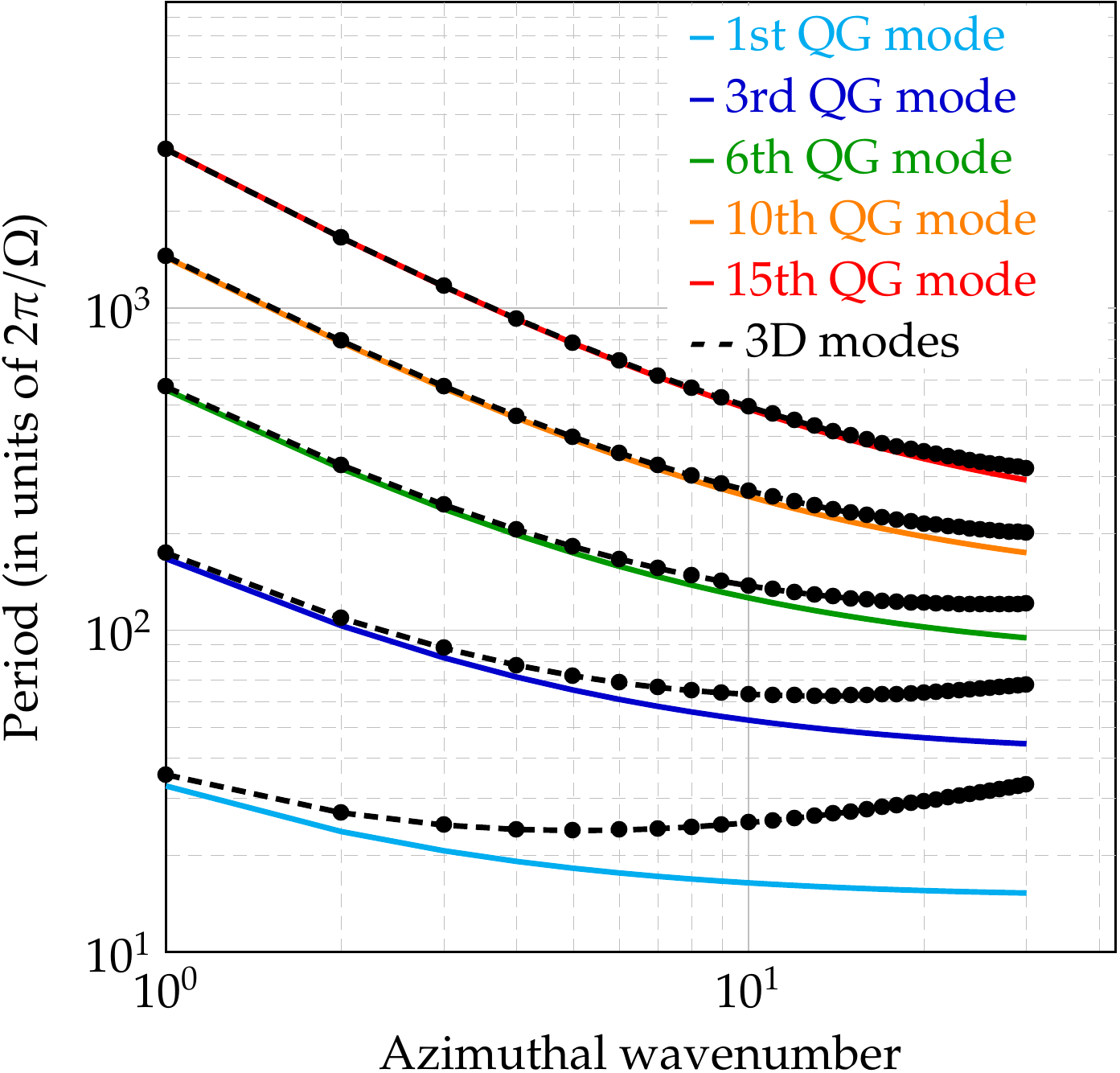}}
\caption{Periods (in units of $2 \pi / \Omega$) of hydrodynamic QG modes in a sphere as a function of azimuthal wavenumber. The coloured lines show the results for quasi-geostrophic modes of increasing complexity (smaller wavelength) in the radial direction.  The corresponding predictions from the \cite{z2001} theory are shown as dashed-bulleted lines.}
\label{comparison3DQG}
\end{minipage}
\end{figure}

Here, we compare the mode frequencies determined from the \citet{z2001} theory \eqref{Zhang_ES_roots} with those computed numerically using our QG model in the absence of any magnetic field, with spherical boundaries and with a vanishingly small inner core (tests showed that provided this is chosen sufficiently small, the inner core has a negligible effect on the modes discussed here).   In this case our QG model reduces to  (\ref{eq:Rossby}) :
\begin{equation}
 \left[ \omega_{qg} \left( 
\frac{1}{s} \partial_s s \partial_s -\frac{m^2}{s^2} -\beta \partial_s 
\right) 
+ 2\frac{m \beta}{s}
\right] \hat{\Psi} 
=
0,
\end{equation}
where, as before, $\beta(s)=H'(s)/H(s)$ and we impose the boundary conditions $\Psi=0$ at $s=1$ and $s=0.01$.  As previously, we expand $\hat{\Psi}$ as a sum of Chebyshev polynomials and impose the equations and boundary conditions using a collocation method.  In this scenario a rather simple real, linear, algebraic eigenvalue problem in $\omega_{qg}$ is obtained.  Results for both $\omega_{si}$ and $\omega_{qg}$ are presented for comparison in Figure \ref{comparison3DQG}, as a function of the azimuthal wavenumber $m$ for five example radial modes.  
 
For the longest period (slowest) modes (with low $m$ but high radial wavenumber) the agreement between the QG numerical model and the predictions of the \citet{z2001} analytic theory is excellent.  This is encouraging since it shows that when the flow disturbances are very close to geostrophic then the QG model performs as desired, giving good agreement with full three-dimensional solutions.  We can also see that for azimuthal wave numbers below $m=10$ the QG model generally performs rather well, especially for disturbances with short wavelength in the radial direction. 

Nonetheless,   there is evidently a poor agreement between our QG model and the three-dimensional theory for the modes with largest radial wavelengths, particularly when the azimuthal wavenumber is very high.  Examination of the associated eigenfunctions shows that, for a given radial mode, as $m$ increases the eigenfunction is increasingly pushed towards the outer boundary; in this region the boundary slope is large and the three-dimensional solutions show more prominent departures from axial invariance that are incompatible with the QG assumption, compare for  example Figures 1 and 2 from \citet{b2005} for $m=6$.  The failure of the QG model manifests itself in terms of an incorrect slope of the dispersion curve at very large $m$, see for example the mode with larges radial length scale (the blue curve in Figure~\ref{comparison3DQG}); physically this implies that the azimuthal group velocity of such modes is incorrect in the QG model. 

We note here in passing that we have presented results from a quasi-geostrophic formulation that includes three-dimensional mass conservation, rather the simpler small-slope approximation that was imposed in early models \citep{s2005,c2009}.  The agreement between the QG model predictions including three-dimensional mass conservation and the \cite{z2001} modes was found to be superior than that for a formulation based on the small-slope approximation.

The results presented here provide evidence that our QG model is a useful tool for study the dynamics of Earth's core, provided most of the flow energy lies at small azimuth wavenumber, and especially if structures also possess short length scales in the cylindrical radial direction.

\end{appendix}
\end{document}